\pdfoutput=1
\documentclass[journal]{IEEEtran}

\usepackage{caption}
\usepackage{graphicx}
\usepackage{amsmath,amssymb} 
\usepackage{color}
\usepackage{multirow}
\usepackage{mathrsfs}
\usepackage{subfigure}
\usepackage{array}
\usepackage{threeparttable}
\usepackage{float}
\ifCLASSINFOpdf

\else

\fi

\begin{document}

\title{A Novel Bi-hemispheric Discrepancy Model for EEG Emotion Recognition}

\author{Yang~Li,~\IEEEmembership{Student Member,~IEEE,}
	Wenming~Zheng$^*$,~\IEEEmembership{Senior Member,~IEEE,}
Lei Wang,~\IEEEmembership{Senior Member,~IEEE,}
Yuan Zong,
Lei Qi,
Zhen~Cui,~\IEEEmembership{Member,~IEEE,}
Tong~Zhang,
Tengfei~Song

\thanks{Yang Li and Tengfei~Song are with the Key Laboratory of Child Development and Learning Science (Ministry of Education), and the Department of Information Science and Engineering, Southeast University, Nanjing, Jiangsu, 210096, China. \protect}
\thanks{Wenming Zheng and Yuan Zong are with the Key Laboratory of Child Development and Learning Science (Ministry of Education), School of Biological Sciences and Medical Engineering,
Southeast University, Nanjing, Jiangsu, 210096, China.\protect \it{($^*$Corresponding author: Wenming Zheng (E-mail: wenming\_zheng@seu.edu.cn).)}}
\thanks{Lei Wang is with the School of Computing and Information Technology, University of Wollongong, NSW, 2500, Australia.\protect}
\thanks{Lei Qi is with the State Key Laboratory for Novel Software Technology, Nanjing University, Nanjing, Jiangsu, 210096, China.}
\thanks{Tong~Zhang and Zhen Cui are with the School of Computer Science and Engineering, Nanjing University of Science and Technology, Nanjing, Jiangsu, 210096, China.\protect}
}

\markboth{}
{Shell \MakeLowercase{\textit{et al.}}: Bare Demo of IEEEtran.cls for Journals}

\maketitle
\begin{abstract}
	The neuroscience study~\cite{dimond1976differing} has revealed the discrepancy of emotion expression between left and right hemispheres of human brain. Inspired by this study, in this paper, we propose a novel bi-hemispheric discrepancy model (BiHDM) to learn the asymmetric differences between two hemispheres for electroencephalograph (EEG) emotion recognition. Concretely, we first employ four directed recurrent neural networks (RNNs) based on two spatial orientations to traverse electrode signals on two separate brain regions, which enables the model to obtain the deep representations of all the EEG electrodes' signals while keeping the intrinsic spatial dependence. Then we design a pairwise subnetwork to capture the discrepancy information between two hemispheres and extract higher-level features for final classification. Besides, in order to reduce the domain shift between training and testing data, we use a domain discriminator that adversarially induces the overall feature learning module to generate emotion-related but domain-invariant feature, which can further promote EEG emotion recognition. We conduct experiments on three public EEG emotional datasets, and the experiments show that the new state-of-the-art results can be achieved.
\end{abstract}


\IEEEpeerreviewmaketitle


\section{Introduction}
Emotion, as a common mental phenomenon, is closely related to our daily life. Although it is easy to sense other people's emotion in human-human interaction, it is still difficult for machines to understand the complicated emotions of human beings~\cite{picard2000affective}. As the first step to make machines capture human emotions, emotion recognition has received substantial attention from human-machine-interaction (HMI) and pattern recognition research communities in recent years~\cite{cowie2001emotion,lin2010eeg,zheng2018emotionmeter}.

Human emotional expressions are mostly based on verbal behavior methods (e.g., speech), and nonverbal behavior methods (e.g., facial expression). Thus, a large body of literature concentrates on learning the emotional components from speech and facial expression data. However, from the viewpoint of neuroscience, human’s emotion originates from a variety of brain cortex regions, such as the orbital frontal cortex, ventral medial prefrontal cortex, and amygdala~\cite{britton2006neural}, which provides us a potential approach to decode emotion by recording the continuous human brain activity signals over these brain regions. For example, by placing the EEG electrodes on the scalp, we can record the neural activities in the brain, which can be used to recognize human emotions.

Most existing EEG emotion recognition methods focus on two fundamental challenges. One is how to extract discriminative features related to emotions. Typically, EEG features can be extracted from the time domain, frequency domain, and time-frequency domain. In~\cite{jenke2014feature}, Jenke et al. evaluated all the existing features by using machine learning techniques on a self-recorded dataset. The other challenge is how to classify the features correctly. Many EEG emotion recognition models and methods have been proposed over the past years~\cite{alarcao2017emotions,garcia2019review}. For example, Zheng et al.~\cite{zheng2016Multichannel} proposed a group sparse canonical correlation analysis method for simultaneous EEG channel selection and emotion recognition. Li et al.~\cite{li2019eeg} fused the information propagation patterns and activation difference in the brain to improve the performance of emotional recognition. These techniques have shown excellent performance on some EEG emotional datasets.

Recently, many researchers have attempted to consider the neuroscience findings of emotion as the prior knowledge to extract features or develop models, effectively enhancing the performance of EEG emotion recognition. For example, Hinrikus et al.~\cite{hinrikus2009electroencephalographic} used EEG spectral asymmetry index for the depression detection. It is well realized through neuroscience study that although the anatomy of human brain looks like symmetric, the left and right hemispheres have different responses to emotions. For example, from the view of neuroscience, Dimond et al.~\cite{dimond1976differing}, Davidson et al.~\cite{davidson1990approach}, and Herrington et al.~\cite{herrington2010localization} have studied the asymmetry of emotion expression, and Schwartz et al.~\cite{schwartz1975right}, Wager et al.~\cite{wager2003valence}, and Costanzo et al.\cite{costanzo2015hemispheric} have discussed the emotion lateralization. Furthermore, the literature of EEG emotion recognition has seen the use of asymmetry to classify EEG emotional signal. Lin et al.~\cite{lin2010eeg} investigated the relationships between emotional states and brain activities, and extracted power spectrum density, differential asymmetry power, and rational asymmetry power as the features. Motivated by their previous findings of critical brain areas for emotion recognition, Zheng et al.~\cite{zheng2018emotionmeter} selected six symmetrical temporal lobe electrodes as the critical channels for EEG emotion recognition. Li et al.~\cite{li2018bi} separately extracted two brain hemispheric features and achieved the state-of-the-art classification performance. The above researches demonstrate that it is a promising and fruitful way to integrate the unique characteristics of EEG signal into the machine learning algorithms. It will be an interesting and meaningful topic of how to utilize this discrepancy property of two brain hemispheres to improve EEG emotion recognition. 

Thus, in this paper, we propose a novel neural network model BiHDM to learn the bi-hemispheric discrepancy for EEG emotion recognition. BiHDM aims to obtain the deep discrepant features between the left and right hemispheres, which is expected to contain more discriminative information to recognize the EEG emotion signals. To achieve this goal, we need to solve two major problems, i.e., how to extract the features for each hemispheric EEG data and meanwhile measure the difference between them. Unlike other data structures such as skeletal action data, in which the position of each node varies with time, the EEG data consists of several electrodes that are set under the predefined coordinates on the scalp. Hence, to avoid losing this intrinsic graph structural information of EEG data, we can simplify the graph structure learning process by using the horizontal and vertical traversing RNNs, which will construct a complete relationship graph and generate discriminative deep features for all the EEG electrodes. After obtaining these deep features of each electrodes, we can extract the asymmetric discrepancy information between two hemispheres by performing specific pairwise operations for any paired symmetric electrodes. The concrete process is as follows:
\begin{itemize}
	\item [(1)] Firstly, we employ individual two RNN modules to separately scan all spatial electrodes' data on the left and right hemispheres and generate deep feature representations for all the EEG electrodes. Herein, when the RNN module traverses the spatial regions, it will walk under two predefined stack strategies determined with respect to the horizontal and vertical direction streams;
	\item [(2)] In each stream, we will obtain the deep features of all the electrodes, and perform specific pairwise operations for the paired electrodes. Herein, the rule of identifying the paired electrodes refers to the symmetric locations on the brain scalp, and the pairwise operations include subtraction, division, and inner product. These operations will model the discrepancy information from different aspects. Subsequently, another RNN summarizes all the asymmetric discrepancy information and produces a global deep representation in each directional stream;
	\item [(3)] Finally, we integrate the global features from horizon and vertical streams with learnable linear transformation matrices and use a classifier to map this representation into the label space. Considering the tremendous data distribution shift of EEG emotional signal, especially in the case of subject-independent task where the source (training) and target (testing) data come from different subjects, we leverage a domain discriminator that works cooperatively with the classifier to encourage the emotion-related but domain-invariant data representation appeared.
\end{itemize}

To the best of our knowledge, this is the first work to integrate the electrodes' discrepancy relation on two hemispheres into deep learning models to improve EEG emotion recognition. The experimental results verify the discrimination and effectiveness of this differential information between the left and right hemispheres for EEG emotion recognition.

The remainder of this paper is organized as follows: In section~\ref{Sec: The proposed method}, we specify the method of BiHDM as well as its application to EEG emotion recognition. In section~\ref{Sec: Experiment}, we conduct extensive experiments to evaluate the proposed method for EEG emotion recognition. In sections~\ref{Sec: Discussion} and~\ref{Sec: Conclusion}, we discuss the paper and conclude it.

\section{The proposed model for EEG emotion recognition}
\label{Sec: The proposed method}
\begin{figure*}[htb]
	\centering
	\includegraphics[width=1.75\columnwidth]{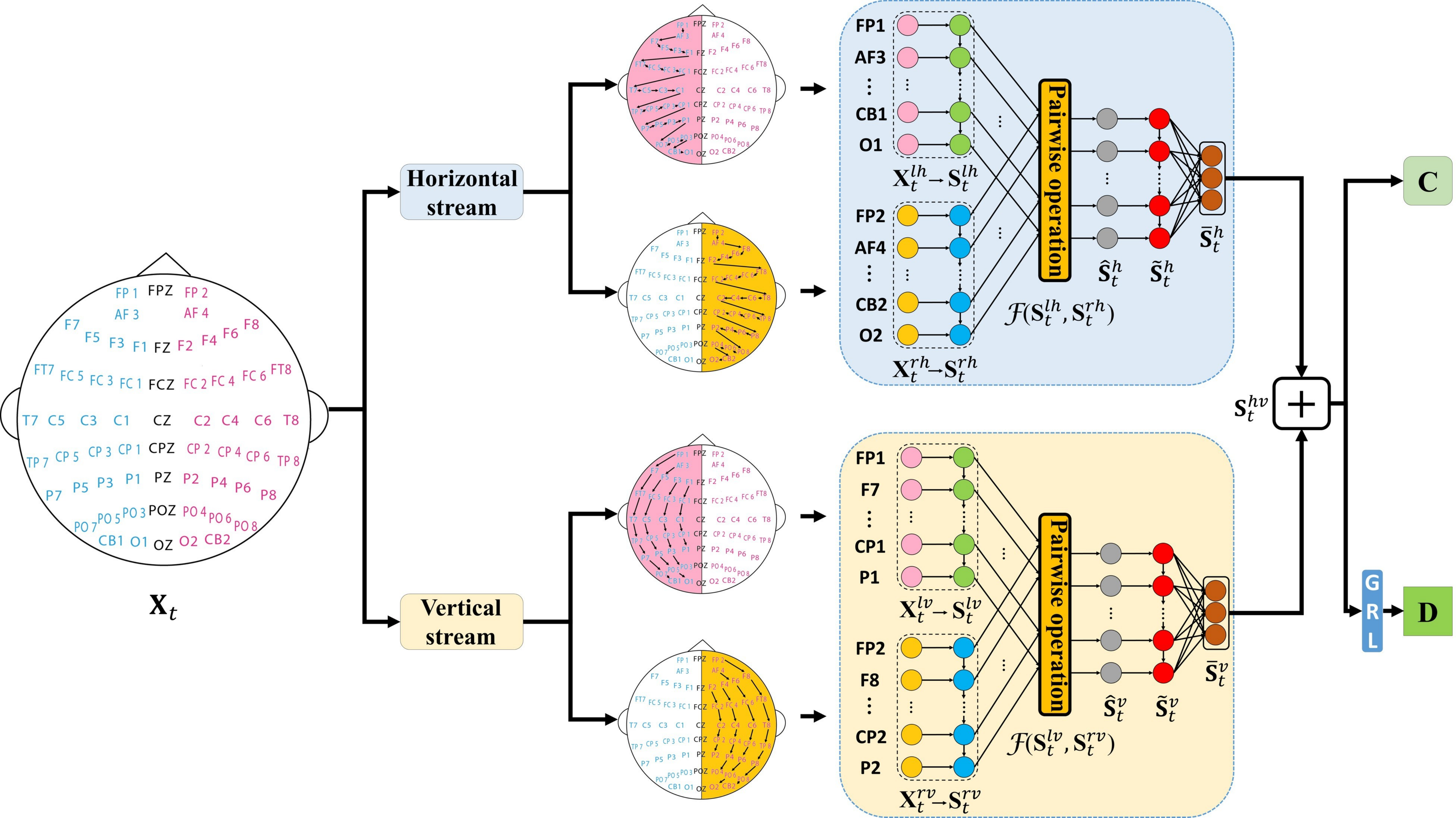} \\
	\caption{The framework of BiHDM. BiHDM consists of four RNN modules to capture each hemispheric EEG electrodes' information from horizontal and vertical streams. Then all the electrodes' data representations interact and construct the final vector for the classifier and discriminator.}
	\label{Fig: BiHDM framework}
\end{figure*}

\subsection{The BiHDM model}
To specify the proposed method clearly, we illustrate the framework of the BiHDM model in Fig.~\ref{Fig: BiHDM framework}. Its goal is to capture the asymmetric differential information between two hemispheres. We adopt three steps to achieve this goal. First, we obtain the deep representations of all the electrodes' data. Subsequently, we characterize the relationship between the identified paired electrodes on two hemispheres, and generate a more discriminative and higher-level discrepancy feature for final classification. Third, we leverage a classifier and a discriminator to corporately induce the above process to generate the emotion-related but domain-invariant features. The overall process is described as follows.

\subsubsection{Obtaining the deep representation for each electrode}

In BiHDM, we attempt to separately extract the EEG electrodes' deep features on left and right brain hemispheres by using two independent RNN modules. To avoid losing the intrinsic graph structural information of EEG data, for each hemispheric EEG data, we build the RNN module traversing the spatial regions under two predefined stacks, which are determined with respect to horizontal and vertical directions. These two directional RNNs are actually complementary for simplifying the technology to construct a complete relationship graph of electrodes' locations. Concretely, for an EEG sample $\mathbf{X}_t$, it can be split as $\mathbf{X}_t \!=\! [\mathbf{X}_t^l, \mathbf{X}_t^r] \!=\! [\mathbf{x}^l_1,\cdots,\mathbf{x}^l_{\frac{N}{2}}, \mathbf{x}^r_1,\cdots,\mathbf{x}^r_{\frac{N}{2}} ] \!\in\! \mathbb{R}^{d \times N}$, where $\mathbf{X}_t^l$ and $\mathbf{X}_t^r$ denote the EEG electrodes' data on the left and right hemispheres, $d$ is the dimension of each EEG electrode's data and $N$ is the number of electrodes. When modeling spatial dependencies, two graphs, i.e., $\texttt{G}^l \!\!=\!\! \{\texttt{N}^l, \texttt{E}^l\}$ and $\texttt{G}^r \!=\! \{\texttt{N}^r, \texttt{E}^r\}$, are used to separately represent the electrodes' spatial relations on the left and right hemispheres, where $\texttt{N}^l \!=\! \{\mathbf{x}^l_{i}\}$ and $\texttt{N}^r \!=\! \{\mathbf{x}^r_{i}\}, (i \!=\!1,2,\cdots,\frac{N}{2})$ denote the electrode sets, while $\texttt{E}^l \!=\! \{e^l_{ij}\}$ and $\texttt{E}^r \!=\! \{e^r_{ij}\}$ represent the edges between spatially neighboring electrodes. Then we traverse through $\texttt{G}^l$ and $\texttt{G}^r$ separately with a predefined forward evolution sequence so that the input state and the previous states can be defined for an RNN unit. This process can be formulated as
\begin{eqnarray}
\mathbf{s}^l_i \!\!\!\!&=&\!\!\!\! \sigma(\mathbf{U}^l\mathbf{x}^l_i +  \sum\nolimits_{j=1}^{N/2} e^l_{ij}\mathbf{V}^l \mathbf{s}^l_j + \mathbf{b}^l) \in \mathbb{R}^{d_l \times 1}, \\
\mathbf{s}^r_i \!\!\!\!&=&\!\!\!\! \sigma(\mathbf{U}^r\mathbf{x}^r_i +  \sum\nolimits_{j=1}^{N/2} e^r_{ij}\mathbf{V}^r \mathbf{s}^r_j + \mathbf{b}^r) \in \mathbb{R}^{d_r \times 1}, 
\end{eqnarray}
\vspace{-0.25cm}
\begin{align}
e^{\cdot}_{ij} = \left\{
\begin{array}{lr}
1, & \textrm{if $\mathbf{x}^{\cdot}_j$ $\in$ $\mathcal{N}$ ($\mathbf{x}^{\cdot}_i$) },\\
0, & \textrm{otherwise},~~~~~~~~~
\end{array}
\right.
\end{align}	
where $\mathbf{s}^l_i$, $\mathbf{s}^r_i$ and $d_l$, $d_r$ are the hidden units and the dimensions of RNN modules on the left and right hemispheres, respectively; $\sigma(\cdot)$ denotes the nonlinear operation such as Sigmoid function; $\{\mathbf{U}^l \!\!\in\!\! \mathbb{R}^{d_l \times d}$, $\mathbf{V}^l \!\!\in\!\! \mathbb{R}^{d_l \times d_l}$, $\mathbf{b}^l \!\!\in\!\! \mathbb{R}^{d_l \times 1}\}$ and $\{\mathbf{U}^r \!\!\in\!\! \mathbb{R}^{d_r \times d}$, $\mathbf{V}^r \!\!\in\!\! \mathbb{R}^{d_r \times d_r}$, $\mathbf{b}^r \!\!\in\!\! \mathbb{R}^{d_r \times 1}\}$ are the learnable transformation matrices of the two hemispheric RNN modules; and $\mathcal{N}$ ($\mathbf{x}^{\cdot}_i$) denotes the set of predecessors of the node $\mathbf{x}^{\cdot}_i$. Here $d_l \!=\! d_r$. As the RNN modules traverse all the nodes in $\texttt{N}^l$ and $\texttt{N}^r$, the obtained hidden states $\mathbf{s}^l_i$ and $\mathbf{s}^r_i$ can be used as the deep features to represent the $i$-th electrode's data on two hemispheres.

Particularly, for the left and right hemispheric RNN modules, they traverse the spatial regions under two predefined horizontal and vertical stacks. Therefore, we will obtain two paired deep feature sets, i.e.,	$(\mathbf{S}_t^{lh}, \mathbf{S}_t^{rh})$ and $(\mathbf{S}_t^{lv}, \mathbf{S}_t^{rv})$, where $\mathbf{S}_t^{lh} \!=\! \{\mathbf{s}^{lh}_i\}\!\in\! \mathbb{R}^{d_l \times (N/2)}$ and $\mathbf{S}_t^{rh} \!=\! \{\mathbf{s}^{rh}_i\} \!\in\! \mathbb{R}^{d_r \times (N/2)}$ represent the left and right hemispheric electrodes' deep features under horizontal direction, while $\mathbf{S}_t^{lv} \!= \!\{\mathbf{s}^{lv}_i\} \!\in\! \mathbb{R}^{d_l \times (N/2)}$ and $\mathbf{S}_t^{rv} \!=\! \{\mathbf{s}^{rv}_i\} \!\in\! \mathbb{R}^{d_r \times (N/2)}$ represent the deep features under vertical direction. So far, we obtain the deep representation of each electrode, which has the emotional discriminative information and keeps the location structural relation.

\subsubsection{Interaction between the paired electrodes on two hemispheres}

After obtaining the deep features of every electrode above, i.e., $(\mathbf{S}_t^{lh}, \mathbf{S}_t^{rh})$ and $(\mathbf{S}_t^{lv}, \mathbf{S}_t^{rv})$, we perform a specific pairwise operation on the paired electrodes referring to the symmetric locations on the brain scalp to identify the asymmetric differential information between two hemispheres. This operation can be expressed as
\begin{eqnarray}
\label{Eq: left differential asymmetric features}
\hat{\mathbf{S}}_t^h =  \mathcal{F}(\mathbf{S}_t^{lh} , \mathbf{S}_t^{rh})\!\!\!\!\! &= \mathcal{F}(\{\mathbf{s}^{lh}_i\} , \{\mathbf{s}^{rh}_i\}) \in \mathbb{R}^{d_p \times (N/2)},\\
\hat{\mathbf{S}}_t^v =  \mathcal{F}(\mathbf{S}_t^{lv} , \mathbf{S}_t^{rv})\!\!\!\!\! &= \mathcal{F}(\{\mathbf{s}^{lv}_i\} , \{\mathbf{s}^{rv}_i\}) \in \mathbb{R}^{d_p \times (N/2)},
\label{Eq: right differential asymmetric features}
\end{eqnarray} 
where $\hat{\mathbf{S}}_t^h \!=\! \{\hat{\mathbf{s}}^h_i\}$ and $\hat{\mathbf{S}}_t^v \!=\! \{\hat{\mathbf{s}}^v_i\}$ are the deep asymmetric differential features, $\mathcal{F}(\cdot)$ denotes the pairwise operation between any two paired electrodes' data representations. Concretely, for any paired data $(\mathbf{s}^{l\cdot}_i, \mathbf{s}^{r\cdot}_i)$, we perform subtraction, division and inner product on it, which can be formulated as
\begin{align}
\label{Eq: pairwise operation}
\mathcal{F}(\{\mathbf{s}^{l\cdot}_i\} , \{\mathbf{s}^{r\cdot}_i\})  ~~~~~~~~~~~~~~~~~~~~~~~~~~~~~~~~~~~~~~~~~~~~~~~~\nonumber\\ =\left\{
\begin{array}{lr}
\!\!\{\mathbf{s}^{l\cdot}_i - \mathbf{s}^{r\cdot}_i\} \in \mathbb{R}^{d_{p_1}\times \frac{N}{2}}, \vspace{0.15cm}\\ 
\!\!\{\mathbf{s}^{l\cdot}_i / \mathbf{s}^{r\cdot}_i\} \in \mathbb{R}^{d_{p_2}\times \frac{N}{2}}, \vspace{0.15cm}\\
\!\!\{s^{l\cdot}_{i,1} \cdot s^{r\cdot}_{i,1} + \cdots + s^{l\cdot}_{i,\frac{N}{2}} \cdot s^{r\cdot}_{i,\frac{N}{2}}\}  \in \mathbb{R}^{d_{p_3}\times \frac{N}{2}},
\end{array}
\right.
\end{align}
where $d_{p_1} = d_{p_2} = d_l$ and $d_{p_3} = 1$.\footnote{Precisely, division operation models the differential information in terms of the relative ratio of magnitude; the (minus) inner product operation models such information in terms of (dis) similarity.}

To further capture the higher-level discrepancy discriminative features, we utilize another RNN module that performs on the obtained differential asymmetric features $\{\hat{\mathbf{s}}^h_i\}$ and $\{\hat{\mathbf{s}}^v_i\}$ from the horizontal and vertical streams. Formally, the operations on them can be written as
\begin{eqnarray}
\label{Eq: left higher-level differential asymmetric features}
\tilde{\mathbf{s}}^h_i \!\!\!\!&=&\!\!\!\! \sigma(\mathbf{U}^h \hat{\mathbf{s}}^h_i + \mathbf{V}^h \tilde{\mathbf{s}}^h_{i-1} + \mathbf{b}^h) \in \mathbb{R}^{d_g \times 1}, \\
\tilde{\mathbf{s}}^v_i \!\!\!\!&=&\!\!\!\! \sigma(\mathbf{U}^v \hat{\mathbf{s}}^v_i + \mathbf{V}^v \tilde{\mathbf{s}}^v_{i-1} + \mathbf{b}^v) \in \mathbb{R}^{d_g \times 1},
\label{Eq: right higher-level differential asymmetric features}
\end{eqnarray} 
where $\{\mathbf{U}^h \!\in\! \mathbb{R}^{d_g \times d_p}$, $\mathbf{V}^h \!\in\! \mathbb{R}^{d_g \times d_g}$, $\mathbf{b}^h \!\in\! \mathbb{R}^{d_g \times 1}\}$ and $\{\mathbf{U}^v \!\in\! \mathbb{R}^{d_g \times d_p}$, $\mathbf{V}^v \!\in\! \mathbb{R}^{d_g \times d_g}$, $\mathbf{b}^v \!\in\! \mathbb{R}^{d_g \times 1}\}$ are the learnable parameter matrices, and $d_g$ is the hidden unit's dimension of the high-level RNN module. Moreover, to automatically detect the salient information related to emotion among these paired differential features, projection matrices are applied to the higher-level discrepancy discriminative features $\{\tilde{\mathbf{s}}^h_i\}$ and $\{\tilde{\mathbf{s}}^v_i\}$ obtained by Eq.~(\ref{Eq: left higher-level differential asymmetric features}) and~(\ref{Eq: right higher-level differential asymmetric features}). Denoting the projection matrices for the horizontal and vertical traversing directions by $\mathbf{W}^h \!=\! [w^h_{ik}]_{(N/2) \times K}$ and $\mathbf{W}^v \!=\! [w^v_{ik}]_{(N/2) \times K}$, the projection can be written as	
\begin{eqnarray}
\bar{\mathbf{s}}^h_k\!=\! \sigma(\sum\nolimits_{i=1}^{N/2} w^h_{ik} \tilde{\mathbf{s}}^h_i+\hat{\mathbf{b}}^h)\in\mathbb{R}^{d_g \times 1}, ~k\!=\!1,2,\cdots,K, \!\\
\bar{\mathbf{s}}^v_k\!=\! \sigma(\sum\nolimits_{i=1}^{N/2} w^v_{ik} \tilde{\mathbf{s}}^v_i+\hat{\mathbf{b}}^v)\in\mathbb{R}^{d_g \times 1}, ~k\!=\!1,2,\cdots,K.
\end{eqnarray} 

Finally, we use two learnable mapping matrices $\mathbf{G}^h \!\in\! \mathbb{R}^{d_o \times d_g}$ and $\mathbf{G}^v \!\in\! \mathbb{R}^{d_o \times d_g}$ to summarize the stimulus $\bar{\mathbf{S}}_t^h \!=\! \{\bar{\mathbf{s}}^h_k\} \!\in\! \mathbb{R}^{d_g \times K}$ and $\bar{\mathbf{S}}_t^v \!=\! \{\bar{\mathbf{s}}^v_k\} \!\in\! \mathbb{R}^{d_g \times K}$ from two directional streams, namely,
\begin{eqnarray}
\mathbf{S}_t^{hv}= \mathbf{G}^h \bar{\mathbf{S}}_t^h  + \mathbf{G}^v \bar{\mathbf{S}}_t^v \in \mathbb{R}^{d_o \times K}.
\end{eqnarray} 
Until now, for an input EEG sample $\mathbf{X}_t$, the output feature $\mathbf{S}_t^{hv}$ is obtained.	

\subsubsection{Discriminative prediction and domain adversarial strategy}

Like most supervised models, we add a supervision term into the network by applying the softmax function to the output feature $\mathbf{S}_t^{hv} \!\!=\! \{\mathbf{s}^{hv}_k\},(k\!=\!1,\cdots,K)$ to predict the class label.

Let $\mathbf{o} = [(\mathbf{s}^{hv}_1)^\mathrm{T}, (\mathbf{s}^{hv}_2)^\mathrm{T}, \cdots, (\mathbf{s}^{hv}_K)^\mathrm{T} ] \in \mathbb{R}^{1 \times Kd_o}$ denotes the output feature vector, then 
\begin{eqnarray}
\mathbf{y} = \mathbf{o}\mathbf{P} + \mathbf{b}^c = \{y_1,y_2,\cdots,y_C\} \in \mathbb{R}^{1 \times C},
\end{eqnarray}  
where $\mathbf{P} \!\in\! \mathbb{R}^{Kd_o \times C}$ and $\mathbf{b}^c \!\in\! \mathbb{R}^{1 \times C}$ are the transform matrices, and $C$ is the number of emotion types.

Finally, the output vector of BiHDM is fed into the softmax layer for emotion classification, which can be written as
\begin{eqnarray}
\label{Eq: classifier}
P(c|\mathbf{X}_t)= \exp(y_c)/\sum\nolimits^{C}_{i=1} \exp(y_i),
\end{eqnarray}  
where $P(c|\mathbf{X}_t)$ denotes the predicted probability that the input sample $\mathbf{X}_t$ belongs to the $c$-th class. As a result, the label $\tilde{l}_t$ of sample $\mathbf{X}_t$ is predicted as 
\begin{eqnarray}
\tilde{l}_t = arg \max_c  P(c|\mathbf{X}_t).
\end{eqnarray}  

Consequently, the loss function of the classifier can be expressed as
\begin{eqnarray}
\label{Eq: loss function classifier}
L_c(\mathbf{X}_t;\theta_f,\theta_c)= \sum_{k=1}^{M_1} \sum_{c=1}^C-\tau (l_t, c)\times \textrm{log} P(c|\mathbf{X}_t),
\end{eqnarray}
\vspace{-0.25cm}
\begin{align}
\tau(l_t, c)=\left\{
\begin{array}{lr}
1, & \textrm{if $l_t$ =}~c,~\\
0, & \textrm{otherwise}.
\end{array}
\right.
\end{align}
Here $\theta_f$ and $\theta_c$ denote the learnable parameters of the feature extraction module and the classifier, while $l_t$ and $M_1$ are the ground-truth label of sample $\mathbf{X}_t$ and the number of training samples. By minimizing the above loss function, discriminative features could be extracted for emotion recognition.

To align the feature distributions between source and target domains, we adopt the domain adversarial strategy by adding a discriminator into the network. It works cooperatively with the classifier to induce the feature extraction process to generate emotion-distinguishable but domain-invariant features.

Concretely, we predefine the source domain label set $D_S=\{0,0,\cdots,0\}\!\in\!\mathbb{Z}^{M_1\times 1}$ and target domain label set $D_T=\{1,1,\cdots,1\}\!\in\!\mathbb{Z}^{M_2\times 1}$, where $M_2$ is the number of testing samples. Then through maximizing the loss function of the discriminator, which can be denoted as
\begin{eqnarray}
\label{Eq: discriminator loss function}
L_d(\mathbf{X}^S_t,\mathbf{X}^T_{t'};\theta_f,\theta_d) \!\!\!\!\!\!\!\!\!\!\!\!\!\!\!\!\!\!\!\!\!\!\!\!\!\!\!\!\!\!\!\!\!\!\!\!\!\!\!\!\!\!\!\!\!\!\!\!&&\nonumber\\
&&= -\sum\nolimits_{k=1}^{M_1} \textrm{log} P(0|\mathbf{X}^S_t) - \sum\nolimits_{k'=1}^{M_2} \textrm{log} P(1|\mathbf{X}^T_{t'}),~~~~~~~
\end{eqnarray}
the feature extraction process expects to have the ability to generate the data representation to confuse the discriminator to distinguish which domain the input comes from (i.e., the domain-invariant features). Here $\mathbf{X}^S_t$ and $\mathbf{X}^T_{t'}$ denote the $t$-th and $t'$-th sample in the source and target data set respectively, and $\theta_d$ represents the learnable parameter of discriminator. 

\subsection{The optimization of BiHDM}
The overall optimization of BiHDM can be expressed as
\begin{eqnarray}
\label{Eq: overall loss function}
\min L(\mathbf{X};\theta_f,\theta_c,\theta_d) ~~~~~~~~~~~~~~~~~~~~~~~~~~~~~~~~~~~~~~~~~~\nonumber\\
= \min L_c(\mathbf{X}^S;\theta_f,\theta_c) + \max L_d(\mathbf{X}^S,\mathbf{X}^T;\theta_f,\theta_d),
\end{eqnarray}
where $L(\cdot)$ is the loss function of the overall model, and $\mathbf{X}$ denotes the entire data set that consists of the source data set $\mathbf{X}^S$ and target data set $\mathbf{X}^T$, i.e., $\mathbf{X}\!=\![\mathbf{X}^S,\mathbf{X}^T] \!\in\! \mathbb{R}^{d\times N \times (M_1\!+\!M_2)}$.

This max-minimizing loss function will force the parameters of feature extraction module to generate emotion-related but domain-invariant data representation, which benefits for EEG emotion recognition because of the tremendous data distribution shift for EEG emotional signal, especially in the case of subject-independent task where the source and target data come from different subjects.

Specifically, the maximizing problem can be transferred to a minimizing problem by using a gradient reversal layer (GRL)~\cite{ganin2016domain} before the discriminator, which can be optimized by using stochastic gradient
descent (SGD) algorithm~\cite{bottou2010large} easily. GRL acts as an identity transform in the forward-propagation but reverses the gradient sign while performing the back-propagation operation. The overall optimization process follows the rules below
\begin{eqnarray}
\theta_c \!\!\!\!&\leftarrow&\!\!\!\! \theta_c - \alpha \frac{\partial{L_c}}{\partial{\theta_c}},~~
\theta_d \leftarrow \theta_d - \alpha \frac{\partial{L_d}}{\partial{\theta_d}},  \\
\theta_f \!\!\!\!&\leftarrow&\!\!\!\! \theta_f - \alpha (\frac{\partial{L_c}}{\partial{\theta_f}} - \frac{\partial{L_d}}{\partial{\theta_f}}),
\end{eqnarray}
where $\alpha$ is the learning rate. In this way, we can iteratively train the classifier and the discriminator to update the parameters with the similar approach of standard deep learning methods by chain rule.

\section{Experiments}
\label{Sec: Experiment}
\subsection{Setting up}
To evaluate the proposed BiHDM model, in this section, we will conduct experiments on three public EEG emotional datasets. All the three datasets were collected when the participants sat in front of a monitor comfortably and watched emotional video clips. The EEG signals are recorded from 62 electrode channels using ESI NeuroScan with a sampling rate of 1000 Hz. The locations of electrodes are on the basis of the international 10-20 system. Thus in the experiment, we perform the pairwise operation on the 31 paired electrodes based on the symmetric locations on the left and right brain hemispheric scalps. The detailed information of these datasets are described as follows:
\begin{enumerate}
	\item[(1)] \textbf{SEED}~\cite{zheng2015investigating}. SEED dataset contains 15 subjects, and each subject has three sessions. During the experiment, the participants watched three kinds of emotional film clips, i,e, \textit{happy}, \textit{neutral} and \textit{sad}, where each emotion has 5 film clips. Consequently, there are totally 15 trails, and each trail has 185-238 samples for one session of each subject. Then there are totally about 3400 samples in one session;
	
	\item[(2)] \textbf{SEED-IV}\footnote{Note that both SEED-IV and MPED are multi-modal datasets. MPED consists of 30 subjects' EEG data, among which 23 subjects contain multi-modal data. In this experiment, we only use the EEG modal data.\label{Footnote: SEED-IV}}~\cite{zheng2018emotionmeter}. SEED-IV dataset also contains 15 subjects, and each subject has three sessions. But it includes four emotion types with the extra emotion \textit{fear} compared with SEED, and each emotion has 6 film clips. Thus there are totally 24 trails, and each trail has 12-64 samples for one session of each subject. Then there are totally about 830 samples in one session;
	
	\item[(3)] \textbf{MPED}\textsuperscript{\ref {Footnote: SEED-IV}}~\cite{8606087}. MPED dataset contains 30 subjects and each subject has one session. It includes seven refined emotion types, i.e., \textit{joy}, \textit{funny}, \textit{neutral}, \textit{sad}, \textit{fear}, \textit{disgust} and \textit{anger}, and each emotion has 4 film clips. There are totally 28 trails, and each trail has 120 samples. There are totally 3360 samples in one subject.
\end{enumerate}

To evaluate the proposed BiHDM model adequately, we design two kinds of experiments including the subject-dependent and subject-independent ones. We use the released handcrafted features, i.e., the differential entropy (DE) in SEED and SEED-IV, and the Short-Time Fourier Transform (STFT) in MPED, as the input to feed our model. Thus the sizes $d\!\times\! N$ of the input sample $\mathbf{X}_t$ are $5\!\times\!62$, $5\!\times\!62$ and $1\!\times\!62$ for these three datasets, respectively. Moreover, in the experiment, we respectively set the dimension $d_l$ of each electrode's deep representation to 32; the parameters $d_g$ and $K$ of the global high-level feature to 32 and 6; and the dimension $d_o$ of the output feature to 16 without elaborate traversal. Specifically, we implemented BiHDM using TensorFlow on one Nvidia 1080Ti GPU. The learning rate, momentum and weight decay rate are set as 0.003, 0.9 and 0.95 respectively. The network is trained using SGD with batch size of 200. In addition, we adopt the subtraction as the pairwise operation of the BiHDM model in the experiment section, and discuss the other two types of operations in section~\ref{Section: different pairwise operations}.

\subsection{The EEG emotion recognition experiments}
\subsubsection{The subject-dependent experiment}
In this experiment, we adopt the same protocols as~\cite{zheng2015investigating}, \cite{zheng2018emotionmeter} and \cite{8606087}. Namely, for SEED, we use the former nine trails of EEG data per session of each subject as source (training) domain data while using the remaining six trials per session as target (testing) domain data; for SEED-IV, we use the first sixteen trials per session of each subject as the training data, and the last eight trials containing all emotions (each emotion with two trials) as the testing data; for MPED, we use twenty-one trials of EEG data as training data and the rest seven trails consist of seven emotions as testing data for each subject. The mean accuracy (ACC) and standard deviation (STD) are used as the final evaluation metrics for all the subjects in the dataset. 

To validate the superiority of BiHDM, we also conduct the same experiments using twelve methods, including linear support vector machine (SVM)~\cite{suykens1999least}, random forest (RF)~\cite{breiman2001random}, canonical correlation analysis (CCA)~\cite{thompson2005canonical}, group sparse canonical correlation analysis (GSCCA)~\cite{zheng2016Multichannel}, deep believe network (DBN)~\cite{zheng2015investigating}, graph regularization sparse linear regression (GRSLR)~\cite{li2018eeg}, graph convolutional neural network (GCNN)~\cite{defferrard2016convolutional}, dynamical graph convolutional neural network (DGCNN)~\cite{song2018eeg}, domain adversarial neural networks (DANN)~\cite{ganin2016domain}, bi-hemisphere domain adversarial neural network (BiDANN)~\cite{li2018novel}, EmotionMeter~\cite{zheng2018emotionmeter}, and attention-long short-term memory (A-LSTM)~\cite{8606087}. All the methods compared in our paper are the representative ones in the previous studies. We directly take (or reproduce) their results from the literature to ensure a convincing comparison with the proposed method. The results are summarized in Table~\ref{Table: dep}.

\begin{table}[htb]
	\caption{The classification performance for subject-dependent EEG emotion recognition on SEED, SEED-IV and MPED datasets.}
	\centering
	\renewcommand{\arraystretch}{1.3}
	\begin{threeparttable}		
		\begin{tabular}{|c|c|c|c|} 
			\hline
			\multirow{2}{*}{\textbf{Method}} & \multicolumn{3}{c|}{\textbf{ACC / STD (\%)}} \\ \cline{2-4}
			&  SEED            & SEED-IV          &   MPED\\ \hline
			SVM~\cite{suykens1999least}          & 83.99/09.72      & ~56.61/20.05$^*$&~32.39/09.53$^*$\\ \hline
			RF~\cite{breiman2001random}           & 78.46/11.77      & ~50.97/16.22$^*$&~23.83/06.82$^*$\\ \hline
			CCA~\cite{thompson2005canonical}          & 77.63/13.21      & ~54.47/18.48$^*$&~29.08/07.96$^*$\\ \hline
			GSCCA~\cite{zheng2016Multichannel}        & 82.96/09.95      & ~69.08/16.66$^*$&~36.78/07.76$^*$\\ \hline
			DBN~\cite{zheng2015investigating}          & 86.08/08.34      & ~66.77/07.38$^*$&~35.07/11.25$^*$\\ \hline
			GRSLR~\cite{li2018eeg}        & 87.39/08.64      & ~69.32/19.57$^*$&~34.58/08.41$^*$\\ \hline
			GCNN~\cite{defferrard2016convolutional}         & 87.40/09.20      & ~68.34/15.42$^*$&~33.26/06.44$^*$\\ \hline
			DGCNN~\cite{song2018eeg}        & 90.40/08.49      & ~69.88/16.29$^*$&~32.37/06.08$^*$\\ \hline
			DANN~\cite{ganin2016domain}         & 91.36/08.30      & ~63.07/12.66$^*$&~35.04/06.52$^*$\\ \hline
			BiDANN~\cite{li2018novel}       & 92.38/07.04      & ~70.29/12.63$^*$&~37.71/06.04$^*$\\ \hline
			EmotionMeter~\cite{zheng2018emotionmeter}     & $-$              & 70.59/17.01     &$-$             \\ \hline
			A-LSTM~\cite{8606087}       &~88.61/10.16$^*$  & ~69.50/15.65$^*$&~38.99/07.53$^*$\\ \hline
			BiHDM    &\textbf{93.12/06.06} &\textbf{74.35/14.09}&\textbf{40.34/07.59}   \\ \hline
		\end{tabular}
		\begin{tablenotes}[para]
			\footnotesize $*$ indicates the experiment results obtained are based on our own implementation.\\
			$-$ indicates the experiment results are not reported on that dataset.
		\end{tablenotes}
	\end{threeparttable}
	\label{Table: dep}
\end{table}

From Table~\ref{Table: dep}, we can see that the proposed BiHDM model outperforms all the compared methods on all the three public EEG emotional datasets, which verifies the effectiveness of BiHDM. Especially for the result on SEED-IV, the proposed method improves over the state-of-the-art method EmotionMeter by 4$\%$. Besides, we can see that the compared method BiDANN, which also considers the bi-hemispheric asymmetry, achieves a comparable performance. The main difference between BiDANN and BiHDM is that the former adopts two hemispheric local discriminators to separately narrow the left and right hemispheric data distribution gaps in either source or target domain but not directly captures the discrepancy information. In contrast, the latter (i.e., the proposed BiHDM) focuses on constructing model to learn the discrepancy relation between two hemispheres and these differential components are beneficial for emotion recognition. Meanwhile, both the results of BiHDM and BiDANN indicate the importance of considering the difference between the left and right cerebral hemispheric data for EEG emotion recognition.  

To test if the proposed BiHDM is statistically significantly better than the baseline method, paired t-test statistical analysis is conducted at the significant level of 0.05. When the improvement of BiHDM over the method is statistically significant, the results will be underlined in the table. Table~\ref{Table: t-test} shows the t-test statistical analysis results, from which we can see BiHDM is significantly better than the baseline method.
\begin{table}[htb]
	\caption{The t-test statistics analysis between BiHDM and the baseline method at the significance
		level of 0.05. When the improvement of BiHDM over the method is statistically significant, the result will be underlined. }
	\centering
	\renewcommand{\arraystretch}{1.3}
	\begin{threeparttable}		
		\begin{tabular}{|c|ccc|} 
			\hline
			\multirow{2}{*}{\textbf{Method}} & \multicolumn{3}{c|}{\textbf{p-value}} \\ \cline{2-4}
			&  SEED            & SEED-IV          &   MPED\\ \hline
			\multirow{2}{*}{{BiHDM vs. BiDANN}} & 0.0580$^a$ & \underline{0.0344}$^a$ & \underline{0.0488}$^a$\\ 
			           & \underline{0.0451}$^b$ & \underline{0.0188}$^b$& \underline{0.0091}$^b$\\ \hline
		\end{tabular}
		\begin{tablenotes}[para]
			\footnotesize $a$ and $b$ indicate the subject-dependent and independent experiment results respectively.
		\end{tablenotes}
	\end{threeparttable}
	\label{Table: t-test}
\end{table}

Besides,  although the representative methods DANN and BiDANN in Table~\ref{Table: dep} have used the unlabelled testing data to enhance their performance, some compared baseline methods only use the labelled training data to learn the model. To have a fair comparison with them, we follow their setting by taking off the discriminator and only using the labelled training data to conduct the same experiments. The accuracy becomes 91.07$\%$, 72.22$\%$ and 38.55$\%$ on SEED, SEED-IV and MPED datasets, which still achieves comparable performance. This indicates our differential features are indeed more discriminative.

\subsubsection{The subject-independent experiment}
In this experiment, we adopt the leave-one-subject-out (LOSO) cross-validation strategy~\cite{zheng2016personalizing} to evaluate the proposed BiHDM model. LOSO strategy uses the EEG signals of one subject as testing data and the rest subjects' EEG signals as training data. This procedure is repeated such that the EEG signals of each subject will be used as testing data once. Again, the mean accuracy (ACC) and standard deviation (STD) are used as the evaluation metrics. 

In addition, for comparison purpose, we use twelve methods including Kullback-Leibler importance estimation procedure (KLIEP)~\cite{sugiyama2008direct}, unconstrained least-squares importance fitting (ULSIF)~\cite{kanamori2009least}, selective transfer machine (STM)~\cite{chu2017selective}, linear SVM, transfer component analysis (TCA)~\cite{pan2011domain}, transfer component analysis (TCA)~\cite{fernando2013unsupervised}, geodesic flow kernel (GFK)~\cite{gong2012geodesic}, DANN, DGCNN, deep adaptation network (DAN)~\cite{li2018cross}, BiDANN, and A-LSTM, to conduct the same experiments. Note that the distribution gap in the subject-independent task is much larger than the subject-dependent one, so that transfer learning methods always achieve promising performance. Therefore, in the subject-independent task, we include lots of domain adaptation methods in the comparison. By doing so, we can effectively validate the state-of-the-art performance of our method. The results are shown in Table~\ref{Table: ind}.\footnote{Note that the subspace based methods, such as TCA, SA and GFK, are problematic to handle a large amount of EEG data due to the computer memory limitation and computational issue. Therefore, to compare with them we have to randomly select 3000 EEG feature samples from the training data set to train these methods.} 
\begin{table}[htb]
	\caption{The classification performance for subject-independent EEG emotion recognition on SEED, SEED-IV and MPED datasets.}
	\centering
	\renewcommand{\arraystretch}{1.3}
	\begin{threeparttable}		
		\begin{tabular}{|c|c|c|c|}
			\hline
			\multirow{2}{*}{\textbf{Method}} & \multicolumn{3}{c|}{\textbf{ACC / STD (\%)}} \\ \cline{2-4}
			                        &    SEED         & SEED-IV        & MPED\\ \hline
			KLIEP~\cite{sugiyama2008direct}           & 45.71/17.76     &~31.46/09.20$^*$&~18.92/04.54$^*$\\ \hline
            ULSIF~\cite{kanamori2009least}            & 51.18/13.57     &~32.99/11.05$^*$&~19.63/03.81$^*$\\ \hline
            STM~\cite{chu2017selective}               & 51.23/14.82     &~39.39/12.40$^*$&~20.89/03.62$^*$\\ \hline
			SVM~\cite{suykens1999least}               &   56.73/16.29   &~37.99/12.52$^*$&~19.66/03.96$^*$\\ \hline
			TCA~\cite{pan2011domain}                  &   63.64/14.88   &~56.56/13.77$^*$&~19.50/03.61$^*$\\ \hline
			SA~\cite{fernando2013unsupervised}        &   69.00/10.89   &~64.44/09.46$^*$&~20.74/04.17$^*$\\ \hline
			GFK~\cite{gong2012geodesic}               &   71.31/14.09   &~64.38/11.41$^*$&~20.27/04.34$^*$\\ \hline
			A-LSTM~\cite{8606087} 				      &~72.18/10.85$^*$ &~55.03/09.28$^*$&~24.06/04.58$^*$\\ \hline
			DANN~\cite{ganin2016domain}               &   75.08/11.18   &~47.59/10.01$^*$&~22.36/04.37$^*$\\ \hline
			DGCNN~\cite{song2018eeg}                  &   79.95/09.02   &~52.82/09.23$^*$&~25.12/04.20$^*$\\ \hline
			DAN~\cite{li2018cross}                    &   83.81/08.56   & 58.87/08.13    &  $-$           \\ \hline
			BiDANN~\cite{li2018novel}                 &   83.28/09.60   &~65.59/10.39$^*$&~25.86/04.92$^*$\\ \hline
			BiHDM                   &\textbf{85.40/07.53}&\textbf{69.03/08.66}&\textbf{28.27/04.99}   \\ \hline
		\end{tabular}
		\begin{tablenotes}[para]
			\footnotesize $*$ indicates the experiment results obtained are based on our own implementation.\\
			$-$ indicates the experiment results are not reported on that dataset.
		\end{tablenotes}
	\end{threeparttable}
	\label{Table: ind}
\end{table}

From Table~\ref{Table: ind}, it can be clearly seen that the proposed BiHDM method achieves the best performance in the three public datasets, which verifies the effectiveness of BiHDM in dealing with subject-independent EEG emotion recognition. For the three datasets, the improvements on accuracy are 2.2$\%$, 3.5$\%$ and 2.4$\%$, respectively, when compared with the existing state-of-the-art methods. On the other hand, we also perform the paired t-test between BiHDM and the baseline method at the significant level of 0.05 to see whether BiHDM has an improvement of recognition rate. Table~\ref{Table: t-test} shows the t-test statistical analysis results, from which we can see BiHDM is significantly better than the baseline method.

\subsection{Confusion matrix}
\begin{figure}[htb]
	\centering
	\subfigure[SEED]{
		\label{CM:SEED-dep}
		\includegraphics[width=0.375\linewidth]{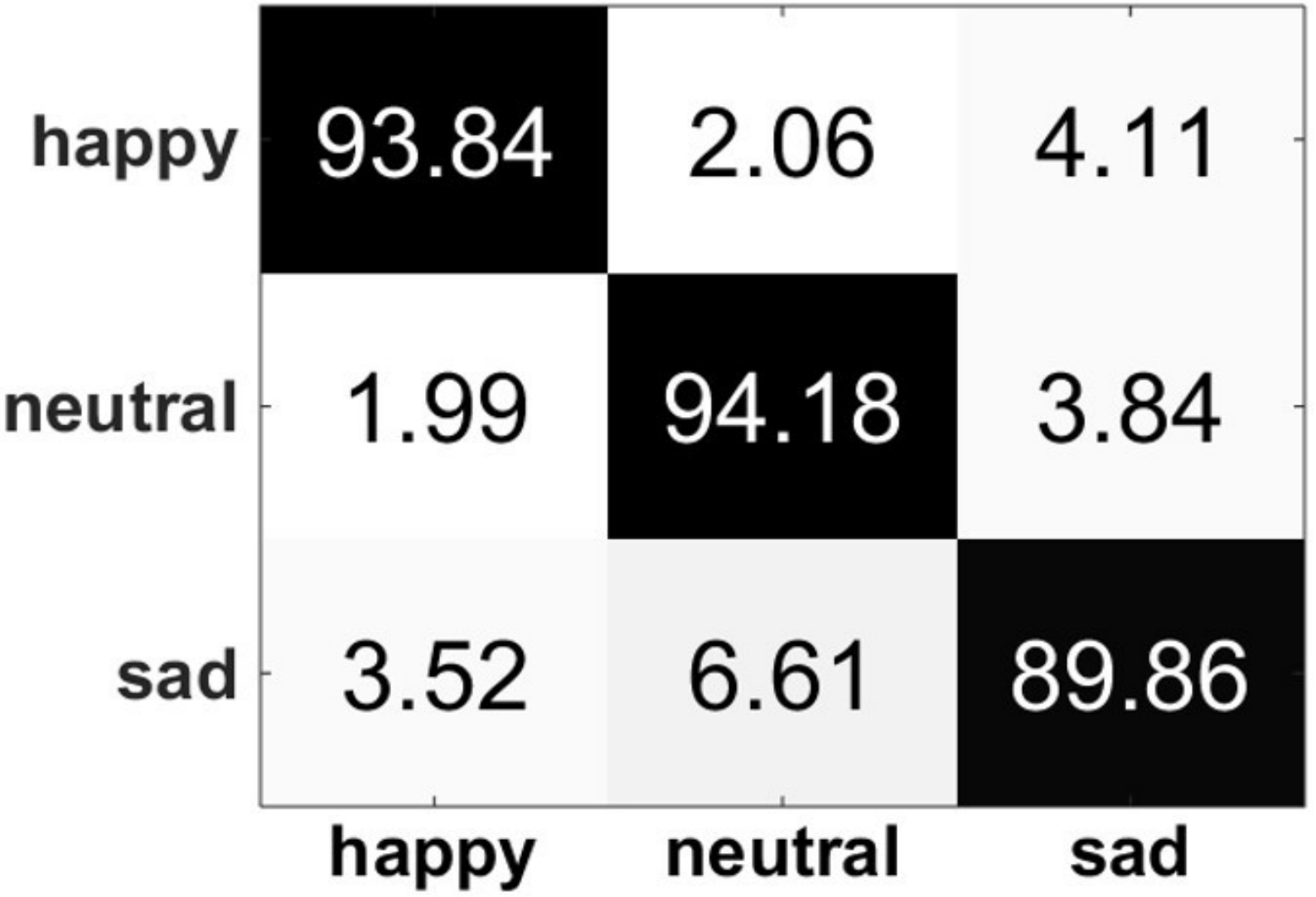}}
	\subfigure[SEED-IV]{
		\label{CM:SEED-IV-dep}
		\includegraphics[width=0.375\linewidth]{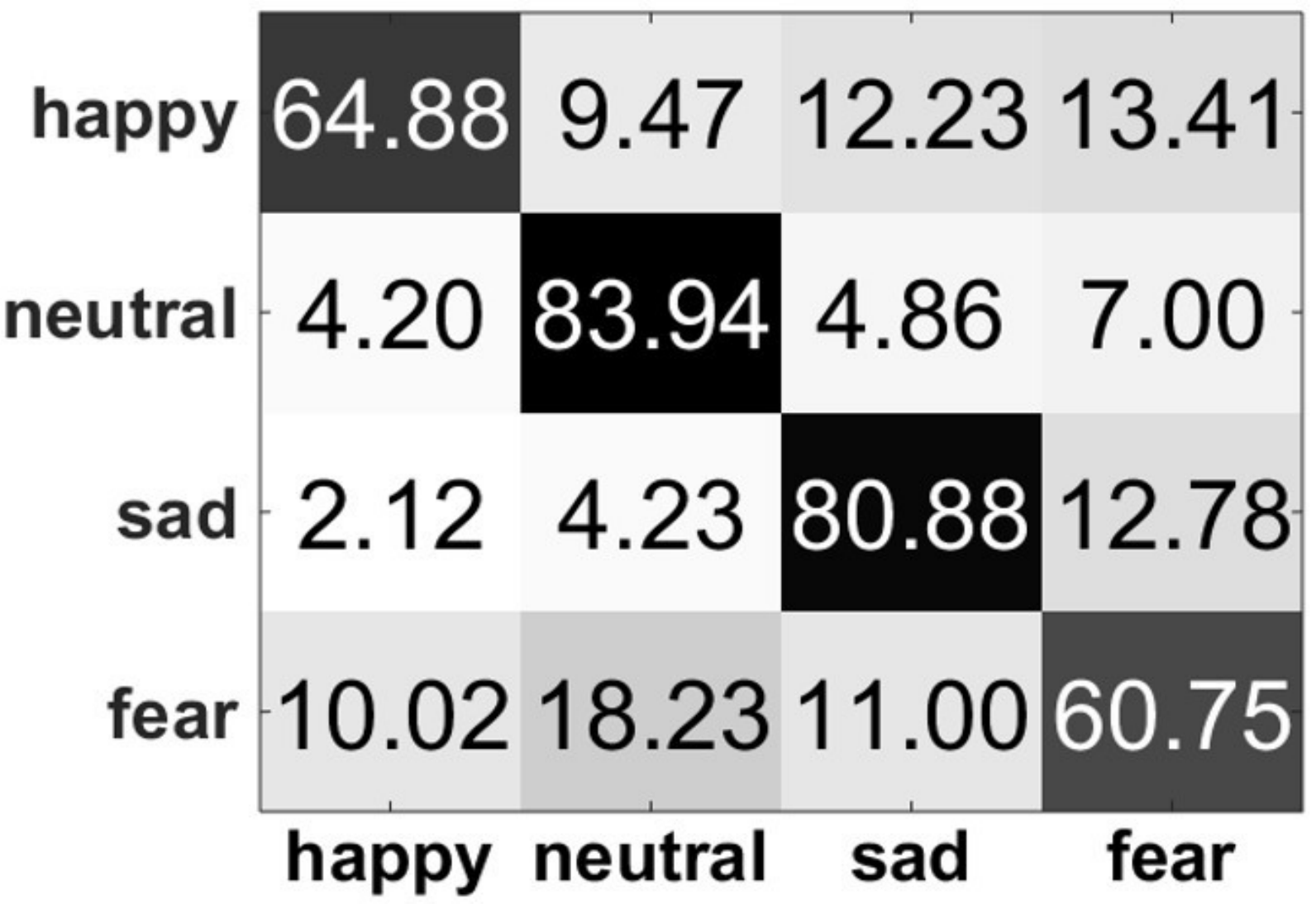}}	
	\subfigure[MPED]{
		\label{CM:MPED-dep}
		\includegraphics[width=0.81\linewidth]{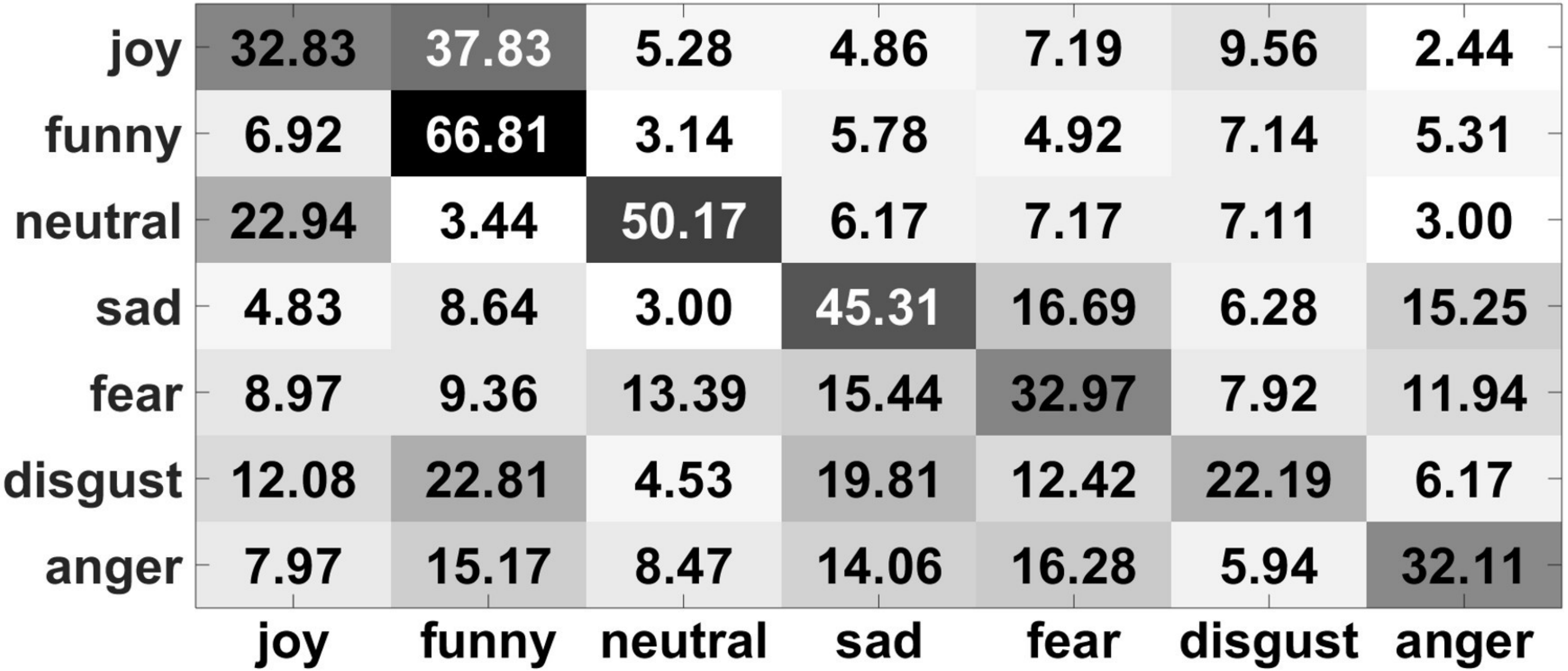}}
	\caption*{(1) Subject-dependent experiment results}
	\subfigure[SEED]{
		\label{CM:SEED-ind}
		\includegraphics[width=0.375\linewidth]{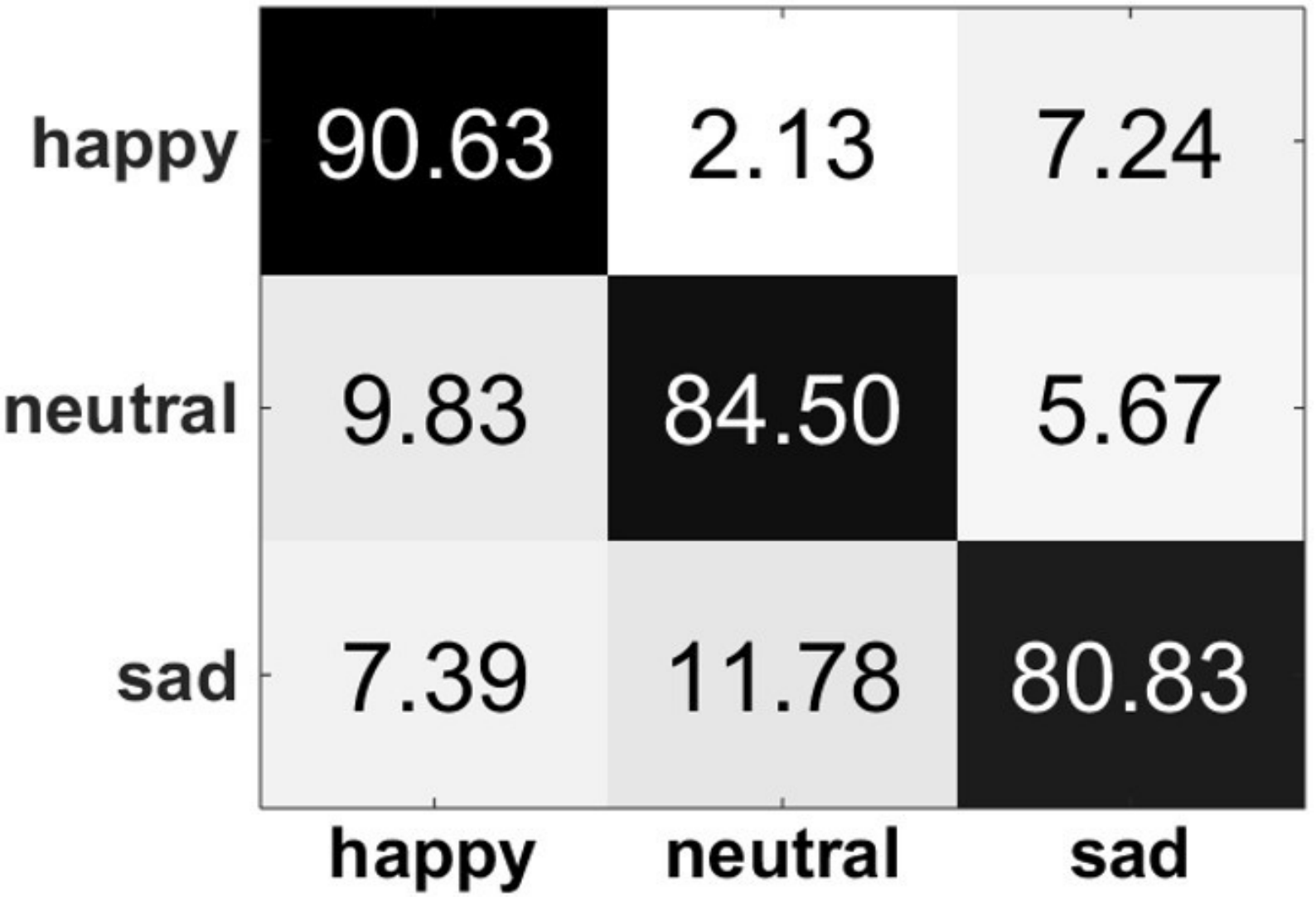}}
	\subfigure[SEED-IV]{
		\label{CM:SEED-IV-ind}	
		\includegraphics[width=0.375\linewidth]{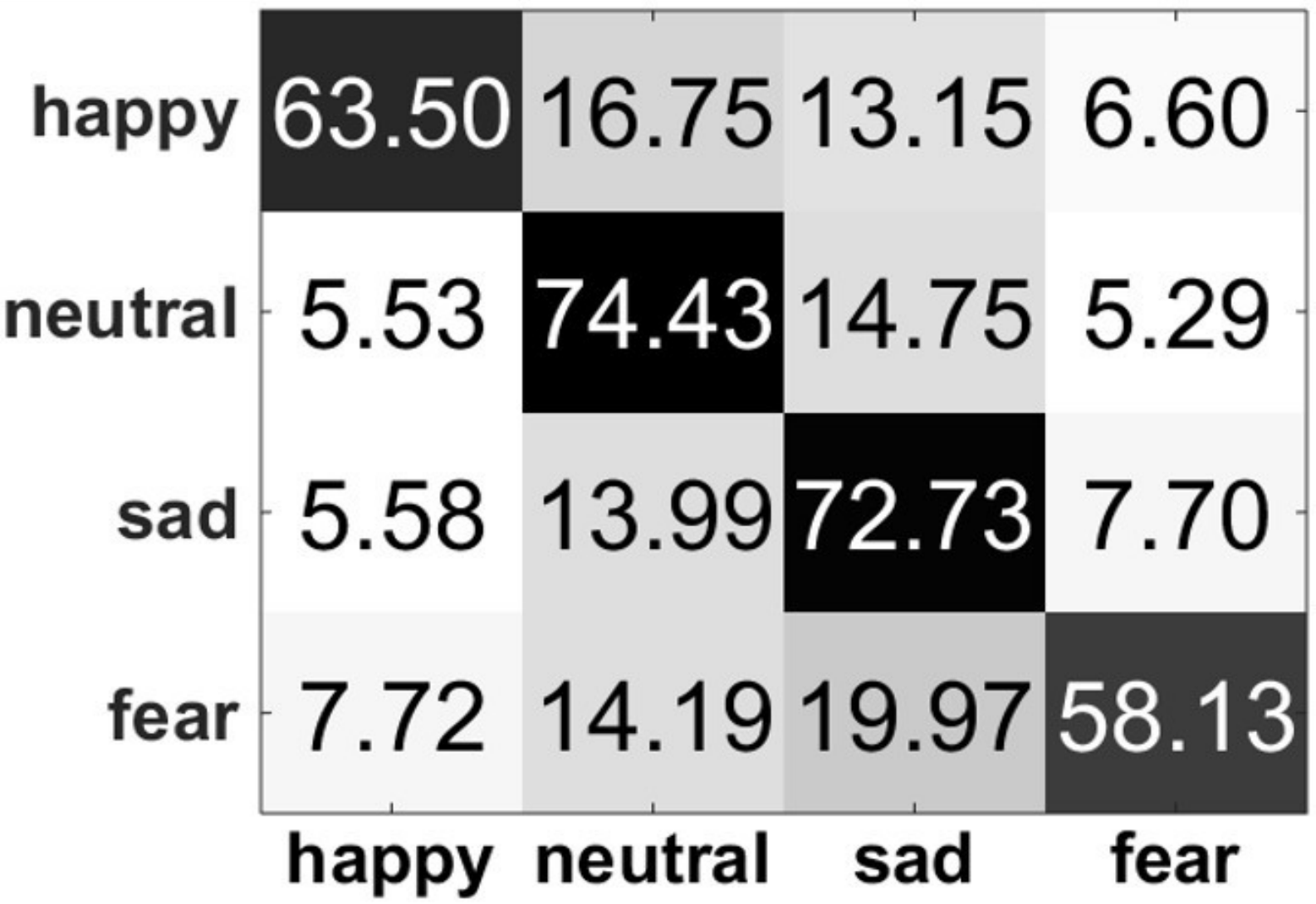}}
	\subfigure[MPED]{
		\label{CM:MPED-ind}
		\includegraphics[width=0.81\linewidth]{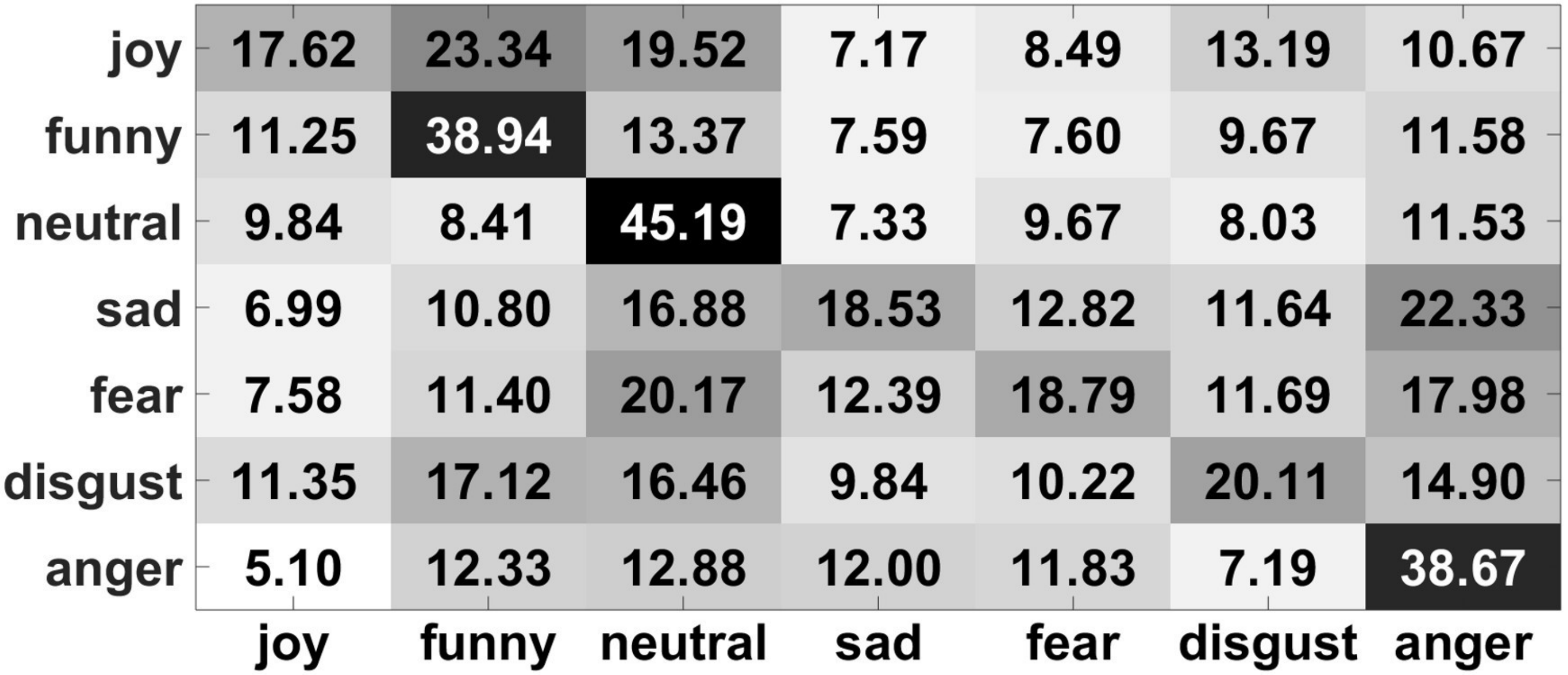}}	
	\caption*{(2) Subject-independent experiment results}
	\caption{The confusion matrices of the experiments.}
	\label{Fig: CM}
\end{figure}

To see the confusions of BiHDM in recognizing different emotions, we depict the confusion matrices of the above two experiments in Fig.~\ref{Fig: CM} from which, we have two observations:
\begin{itemize}
	\item [(1)] From Fig.~\ref{CM:SEED-dep} and Fig.~\ref{CM:SEED-ind} corresponding to the SEED dataset, we can see that the emotions \textit{happy} and \textit{neutral} are much easier to be recognized than \textit{sad}. But comparing the results between these two kinds of experiments, it is easy to see that in the subject-independent experiment, when the training and testing data come from different people, the recognition rates of emotions \textit{neutral} and \textit{sad} will decrease about 10$\%$ and 9$\%$ while the emotion \textit{happy} only decreases 3$\%$. We can also observe the same case from the two confusion matrices of the SEED-IV dataset from Fig.~\ref{CM:SEED-IV-dep} and Fig.~\ref{CM:SEED-IV-ind}. This shows that the emotion \textit{happy} causes more similar brain reflection over different people than \textit{neutral} and \textit{sad};
	\item [(2)] For MPED, which consists of seven emotion types, it is much more complicated than the other two datasets. For the subject-dependent experimental result in Fig.~\ref{CM:MPED-dep}, we can find that the emotions \textit{funny}, \textit{neutral} and \textit{sad} are much easier to be recognized than the other four emotions. However, comparing it with the subject-independent confusion matrix in Fig.~\ref{CM:MPED-ind}, we can see that the recognition rate of \textit{sad} decreases significantly, which is same as the case observed in the above (i.e., point (1)). It is possibly because the pattern of emotion \textit{sad} varies considerably from one subject to another. Moreover, it is interesting to see that the recognition rate of emotion \textit{funny} decreases significantly but the emotion \textit{anger} increases, which may be because the participants share common response to the \textit{anger} emotional videos but have different interpretation about \textit{funny}.
\end{itemize}

\subsection{Different pairwise operations}
\label{Section: different pairwise operations}
\begin{figure}[htb]
	\subfigure[SEED]{
		\includegraphics[width=0.45\linewidth]{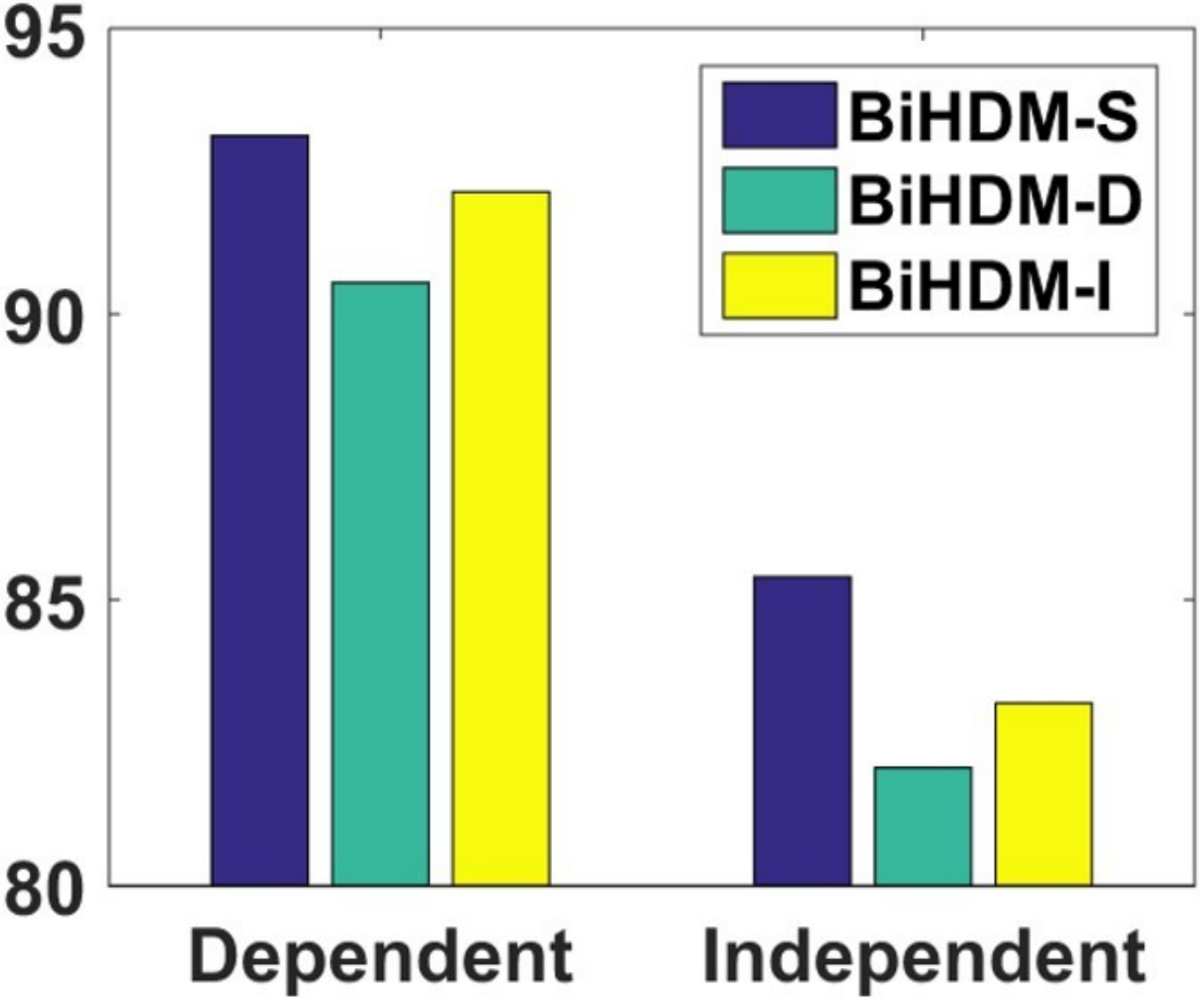}}
	\subfigure[SEED-IV]{
		\includegraphics[width=0.45\linewidth]{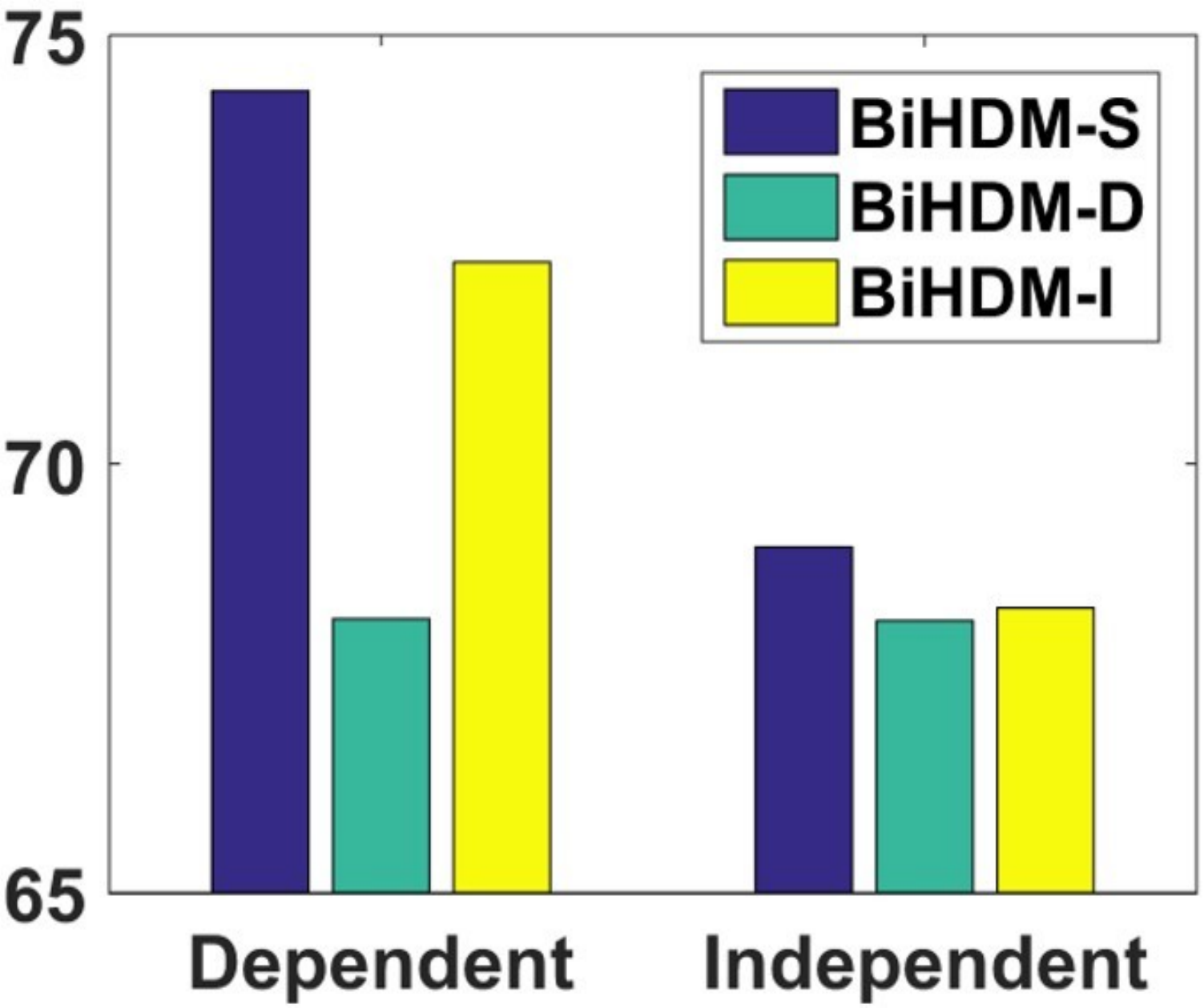}}	
	\subfigure[MPED]{
		\includegraphics[width=0.45\linewidth]{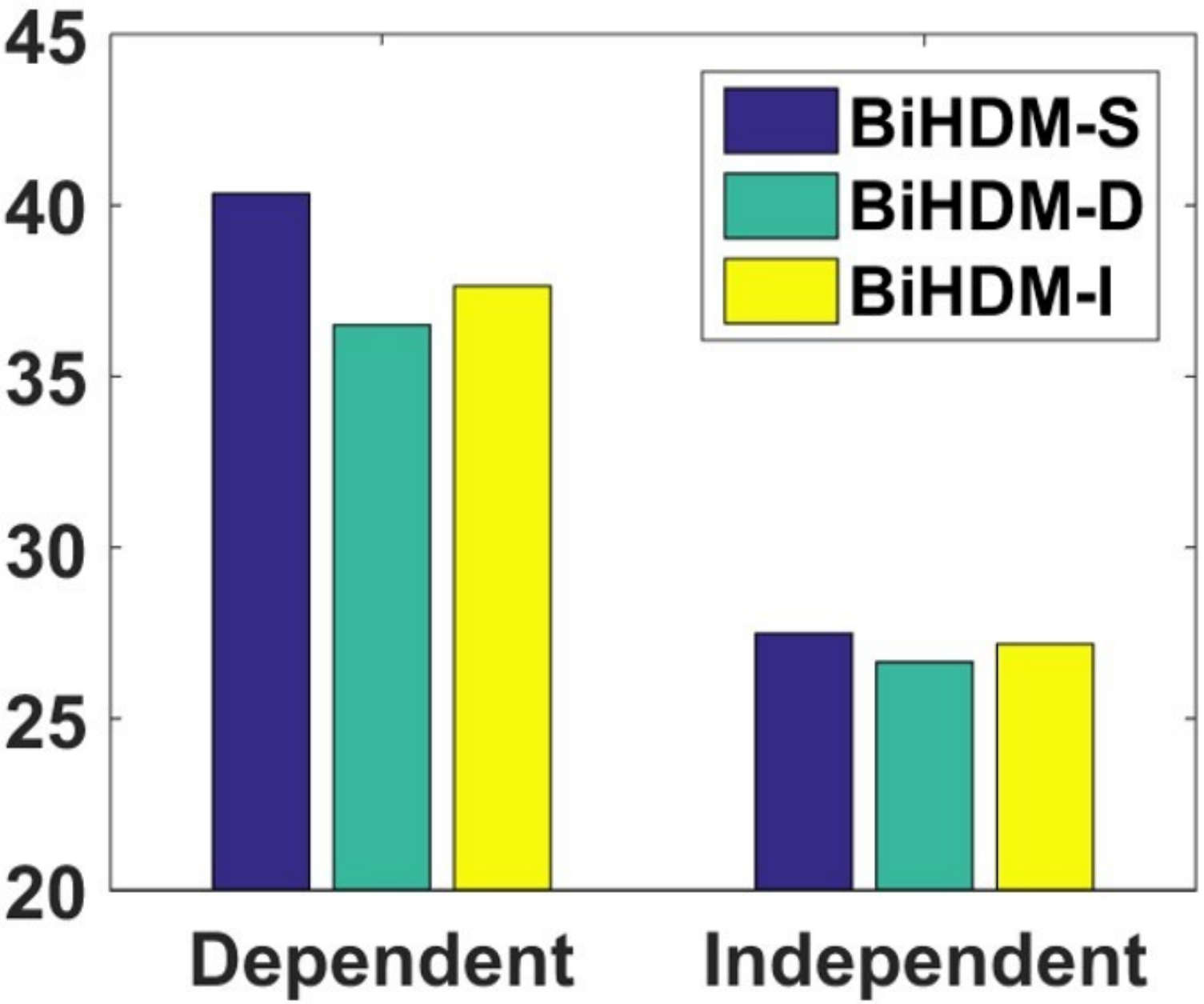}}
	\caption{The experimental results by using BiHDM-S, BiHDM-D and BiHDM-I models on three datasets.}
	\label{Fig: different pairwise operations}
\end{figure}

In this section, we investigate the performance of using different pairwise operations in BiHDM, as shown in Eq.~(\ref{Eq: pairwise operation}). Here, we denote the subtraction, division and inner product variants as BiHDM-S, BiHDM-D, and BiHDM-I, respectively. The results are shown in Fig.~\ref{Fig: different pairwise operations}. As seen, the subtraction operation achieves the best performance among the three pairwise operations. This may be because the subtraction operation directly measure the discrepancy between two hemispheres, whereas the other two operations describe the difference from various aspects. However, both BiHDM-I and BiHDM-D achieve comparable performance compared with the other methods shown in Table~\ref{Table: dep} and~\ref{Table: ind}, which can show the effectiveness of considering the differential information between two cerebral hemispheres. We will explore more pairwise operations such as nonlinear kernel functions in the future work.

To further verify the performance of pairwise operations, we conduct additional experiments for subject-dependent EEG emotion recognition on SEED, SEED-IV and MPED datasets by replacing subtraction with concatenation, and obtain 90.52$\%$, 72.68$\%$ and 37.89$\%$ in accuracy. It is clearly inferior to the proposed subtraction operation (93.12$\%$, 74.35$\%$ and 40.34$\%$). This shows that using pairwise operation to explicitly extract the discrepancy indeed helps EEG emotion recognition.

\section{Discussion}
\label{Sec: Discussion}
\subsection{The activity maps of the paired EEG electrodes}

\begin{figure}[htb]
	\centering
	\subfigure[SEED]{
		\label{Map:SEED-dep}
		\includegraphics[width=0.25\linewidth]{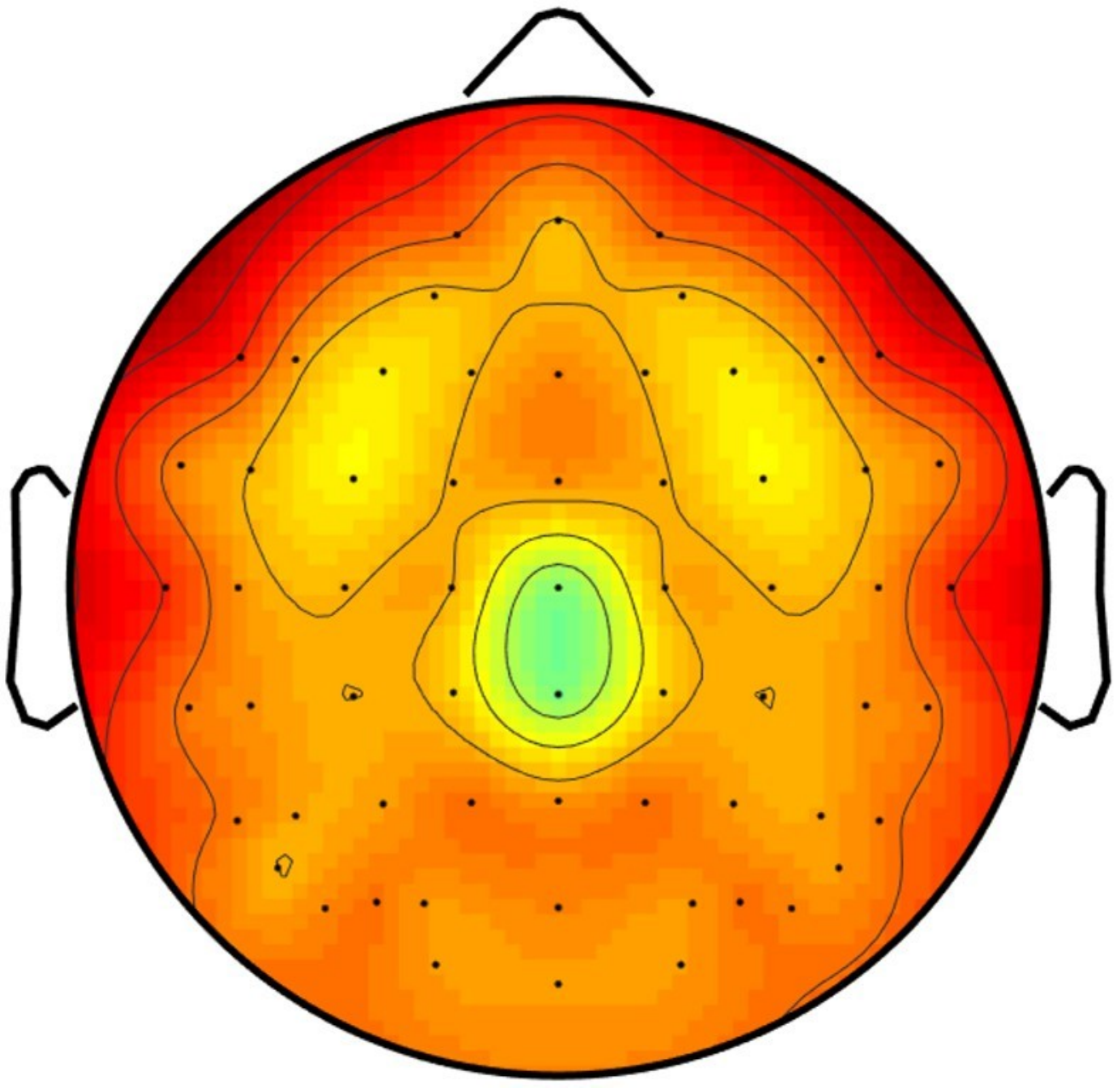}}
	\hspace{0.7cm}
	\subfigure[SEED-IV]{
		\label{Map:SEED-IV-dep}
		\includegraphics[width=0.25\linewidth]{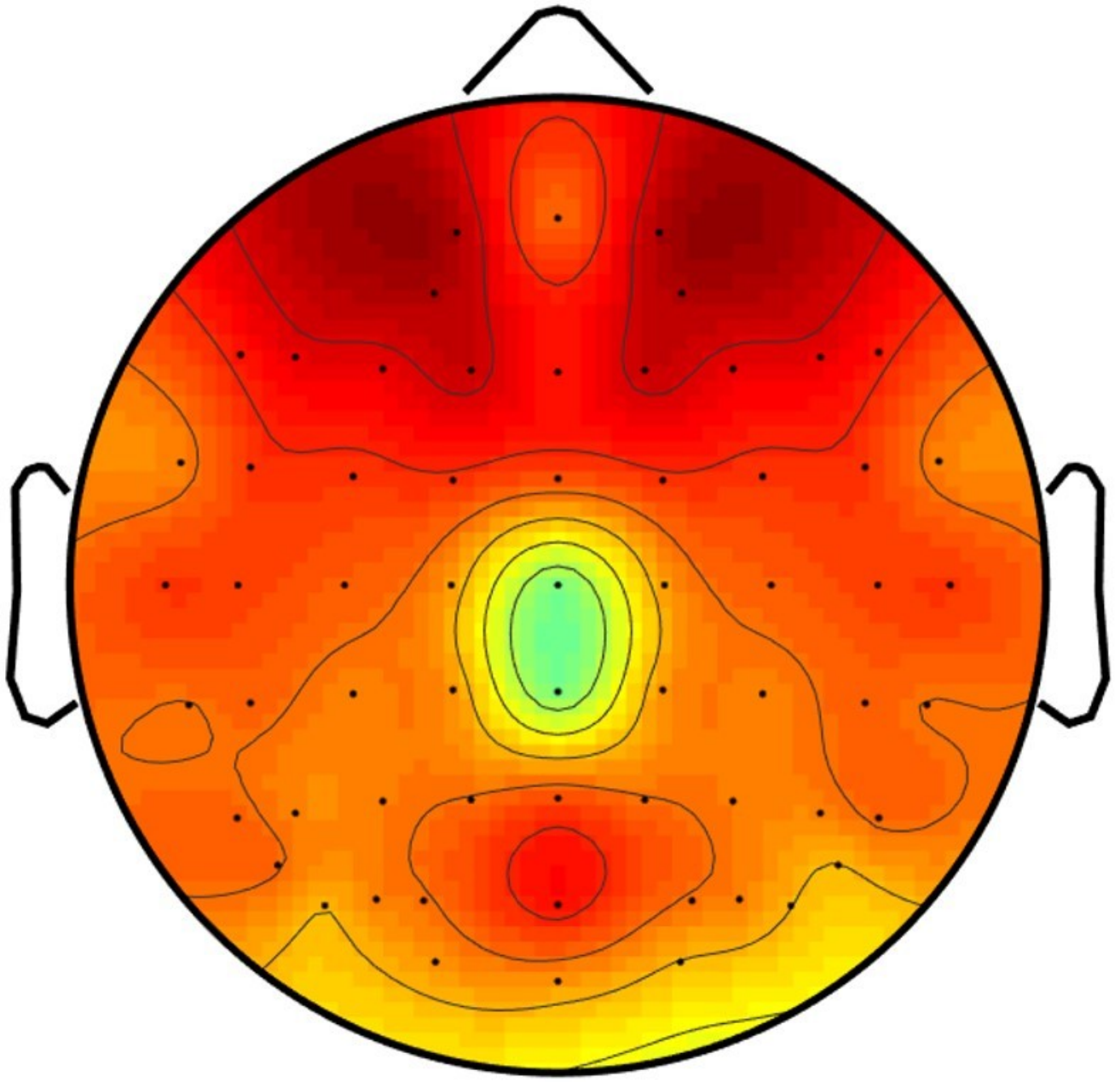}}	
	\hspace{0.7cm}
	\subfigure[MPED]{
		\label{Map:MPED-dep}
		\includegraphics[width=0.25\linewidth]{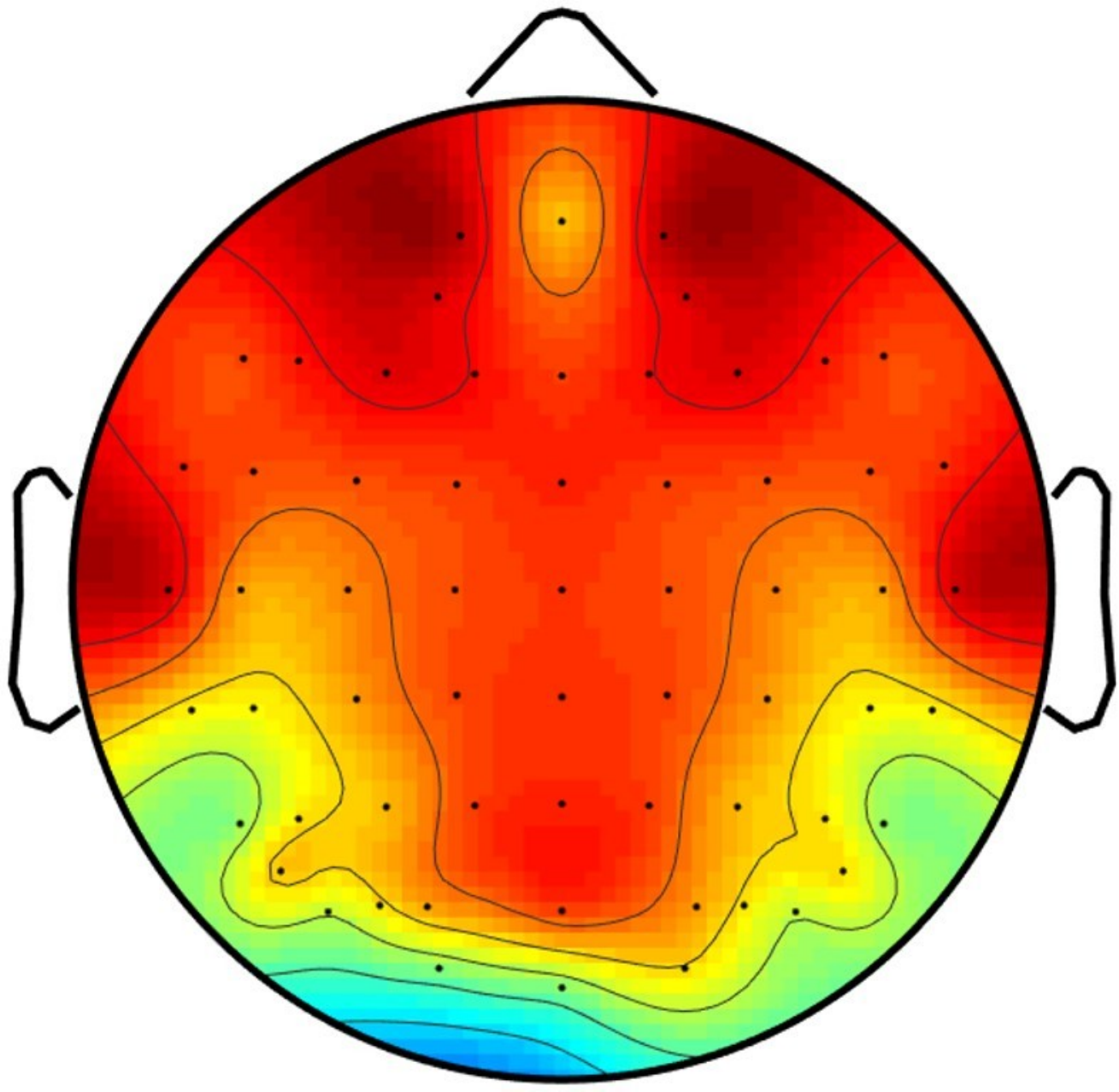}}
	\caption{The EEG electrode activity maps of the subject-dependent experiments. Darker red denotes more significant contribution. Note that these maps are symmetric because it shows the value computed by pairwise operation shared by paired electrodes. (Best Viewed in Color)}
	\label{Fig: The EEG electrode activity maps}
\end{figure}

To explore the contribution of the differential information from various brain areas for emotion expression, we depict the electrode activity maps in Fig.~\ref{Fig: The EEG electrode activity maps}. The contribution is evaluated by computing each column's 2-norm of the asymmetric differential features $\hat{\mathbf{S}}_t^h$ and $\hat{\mathbf{S}}_t^v$ in Eq.~(\ref{Eq: left differential asymmetric features}) and~(\ref{Eq: right differential asymmetric features}) for all the testing data and mapping these values into the corresponding electrodes. The two electrodes in a pair share the same value. From Fig.~\ref{Fig: The EEG electrode activity maps}, we can see that the frontal EEG asymmetry appears to serve as a more important role in emotion recognition for all the three datasets, which is consistent with the cognition observation in biological psychology~\cite{coan2004frontal}. Moreover, for the MPED dataset, which consists of more emotion types, the temporal lobe asymmetry also makes important contribution as the frontal asymmetry. 

\begin{figure*}[htb]
	\centering
	\begin{minipage}[t]{0.42\textwidth}\centering
	\subfigure[Happy]{
		\label{Fig:SEED-dep-happy}
		\includegraphics[width=0.28\linewidth]{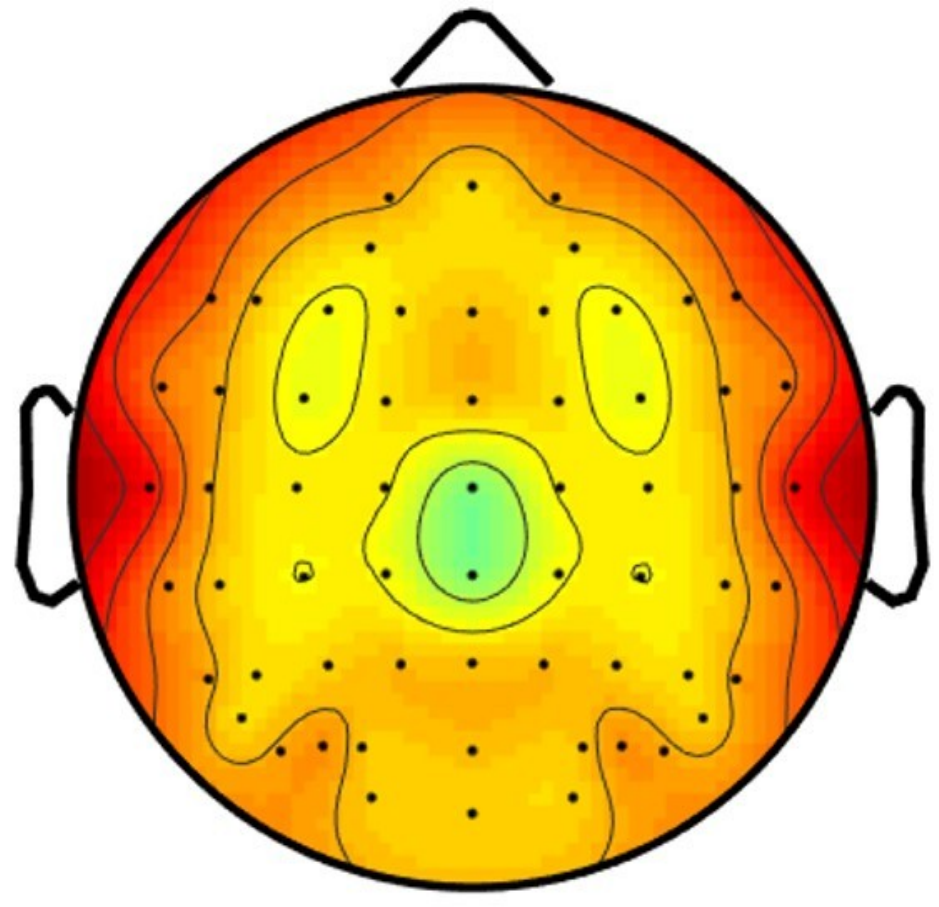}}
	\subfigure[Neutral]{
		\label{Fig:SEED-dep-neutral}
		\includegraphics[width=0.28\linewidth]{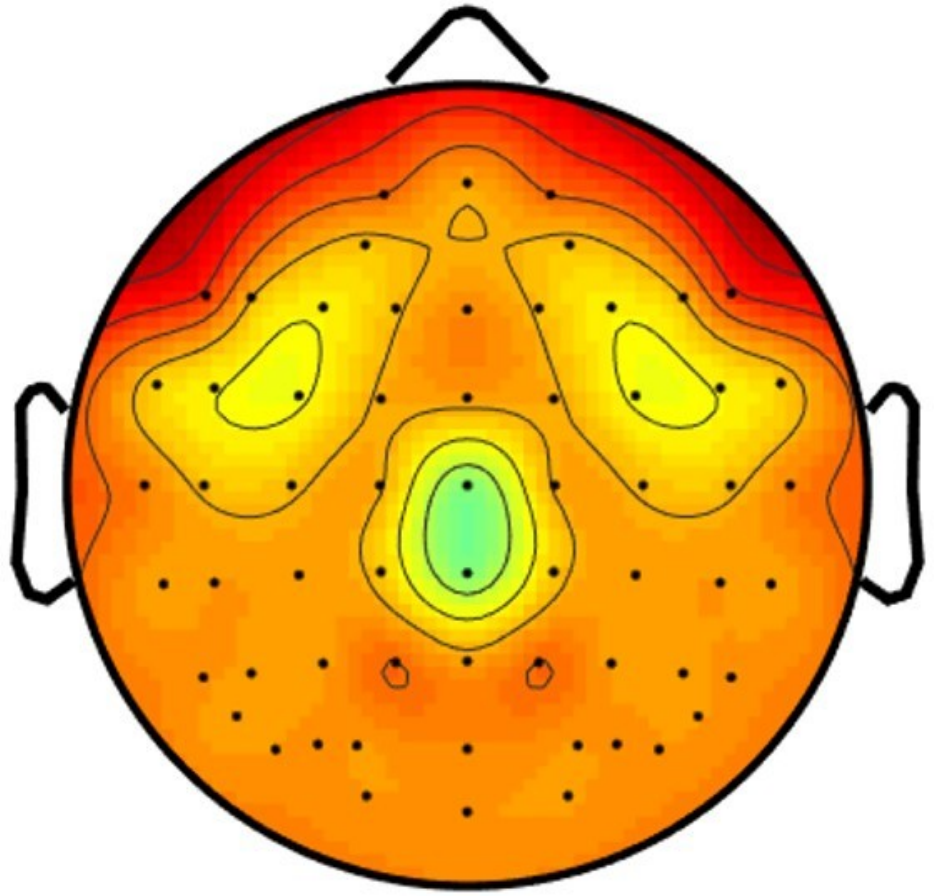}}	
	\subfigure[Sad]{
		\label{Fig:SEED-dep-sad}
		\includegraphics[width=0.28\linewidth]{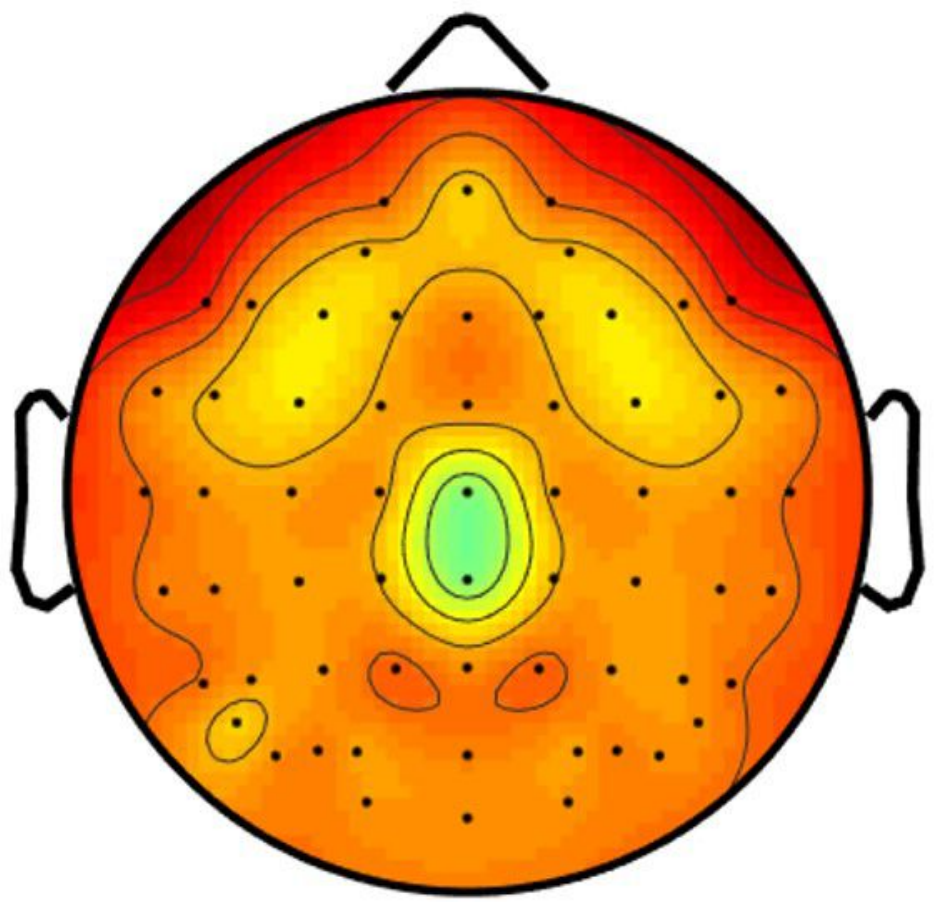}}
	\caption*{(1) SEED}
	\end{minipage}
	\begin{minipage}[t]{0.53\textwidth}\centering
	\subfigure[Happy]{
		\label{Fig:SEED-IV-dep-happy}
		\includegraphics[width=0.22\linewidth]{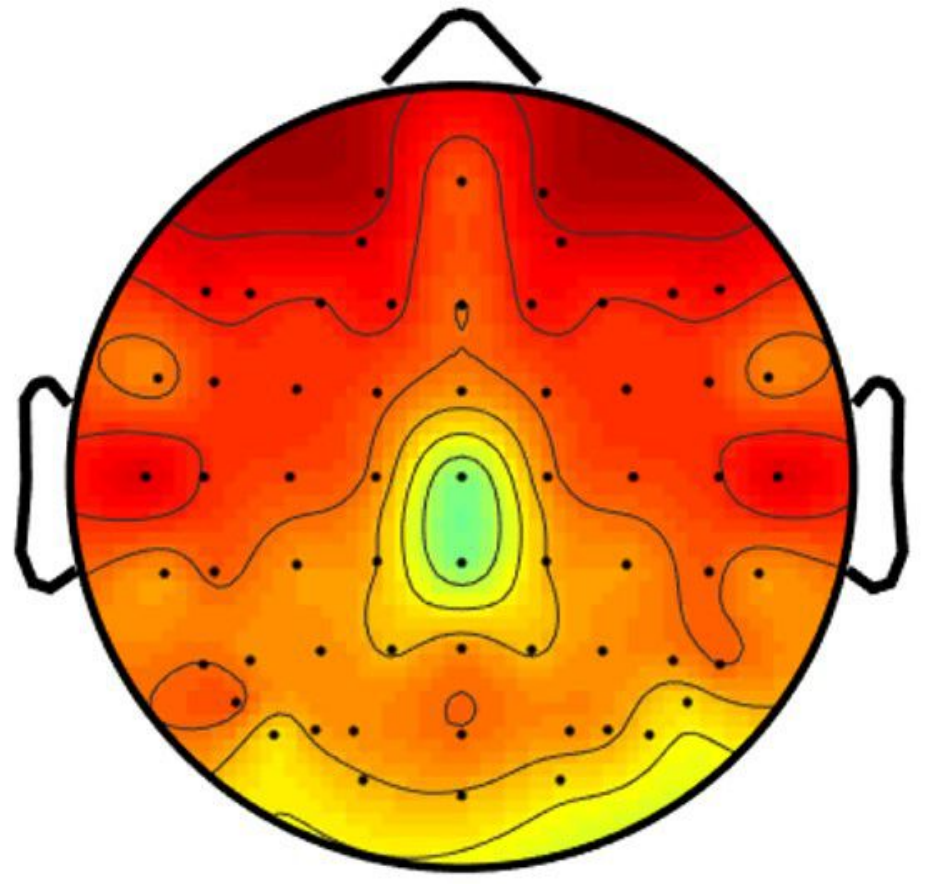}}
	\subfigure[Neutral]{
		\label{Fig:SEED-IV-dep-neutral}
		\includegraphics[width=0.22\linewidth]{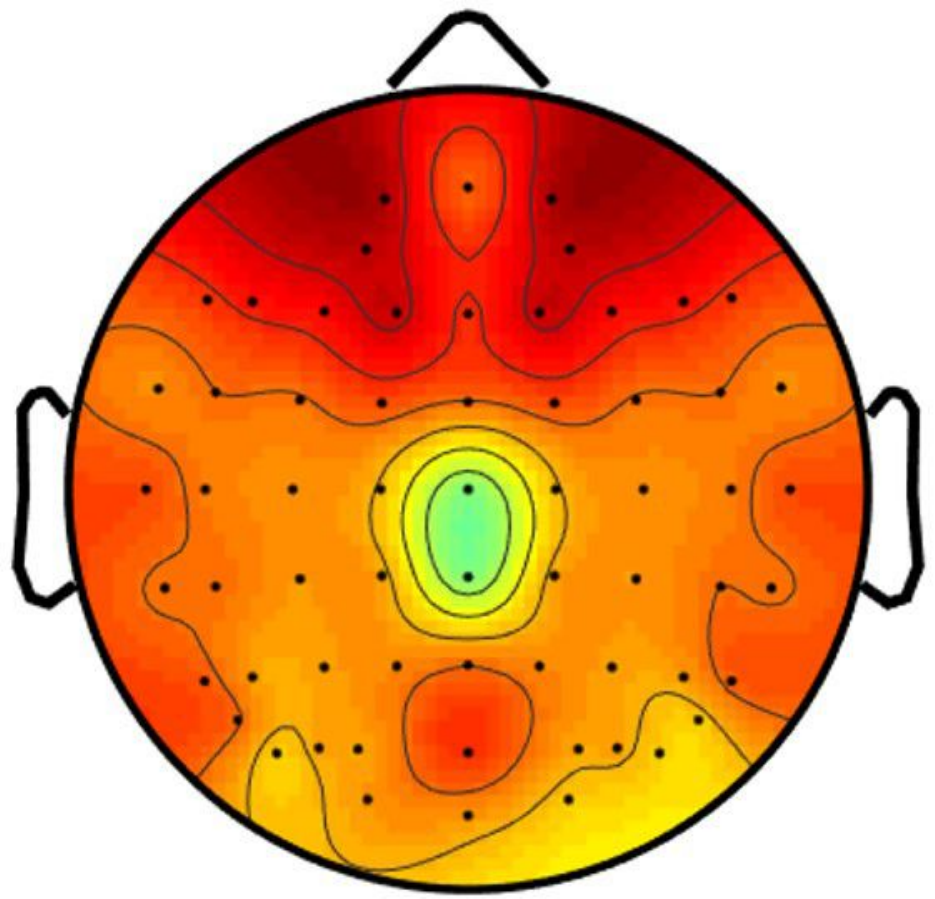}}	
	\subfigure[Sad]{
		\label{Fig:SEED-IV-dep-sad}
		\includegraphics[width=0.22\linewidth]{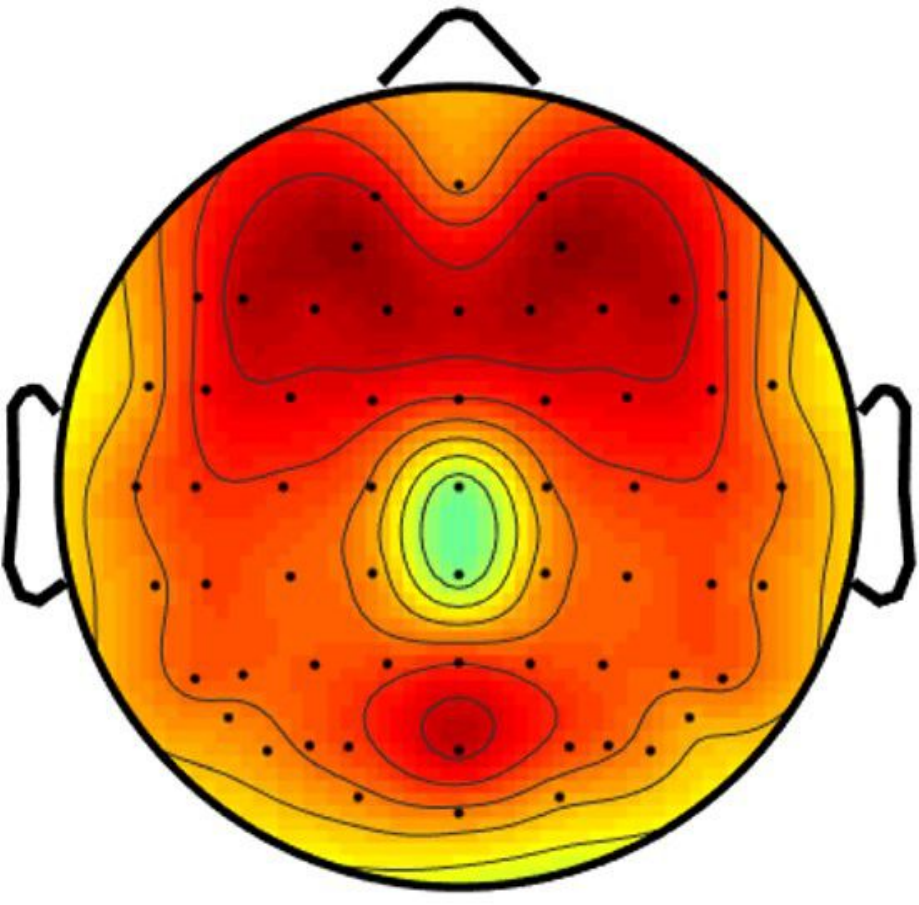}}
	\subfigure[Fear]{
		\label{Fig:SEED-IV-dep-fear}
		\includegraphics[width=0.22\linewidth]{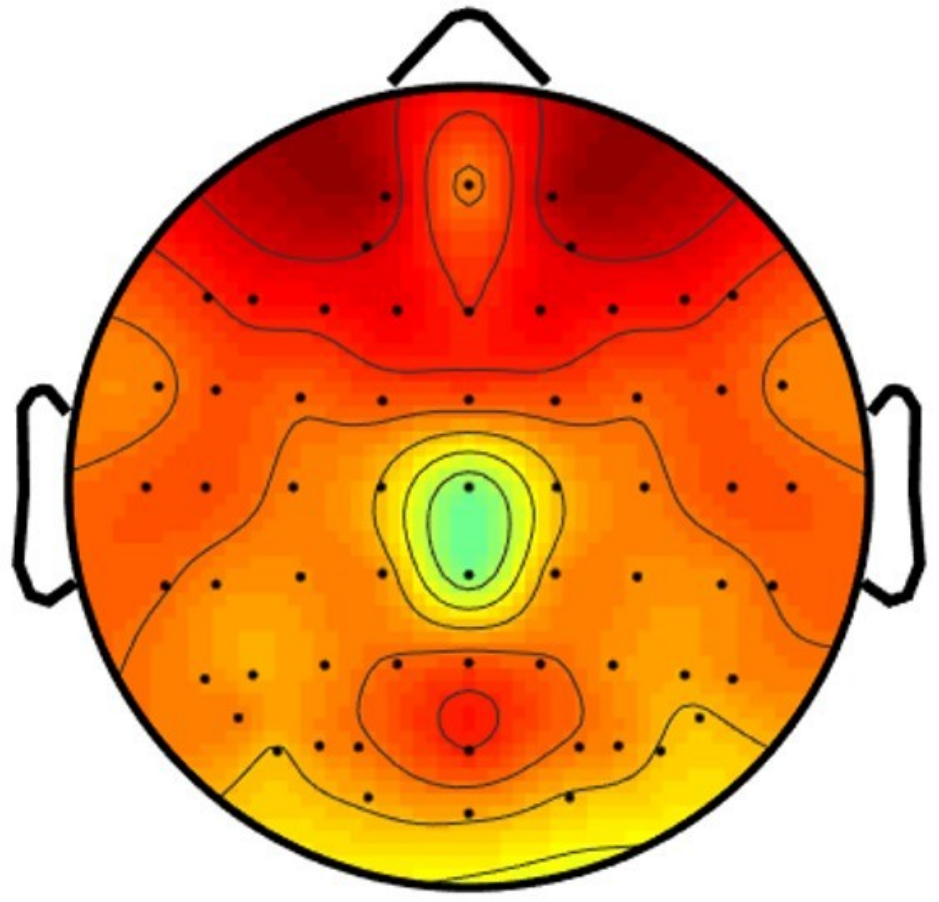}}
	\caption*{(2) SEED-IV}	
	\end{minipage}
	\subfigure[Joy]{
		\label{Fig:MPED-dep-joy}
		\includegraphics[width=0.12\linewidth]{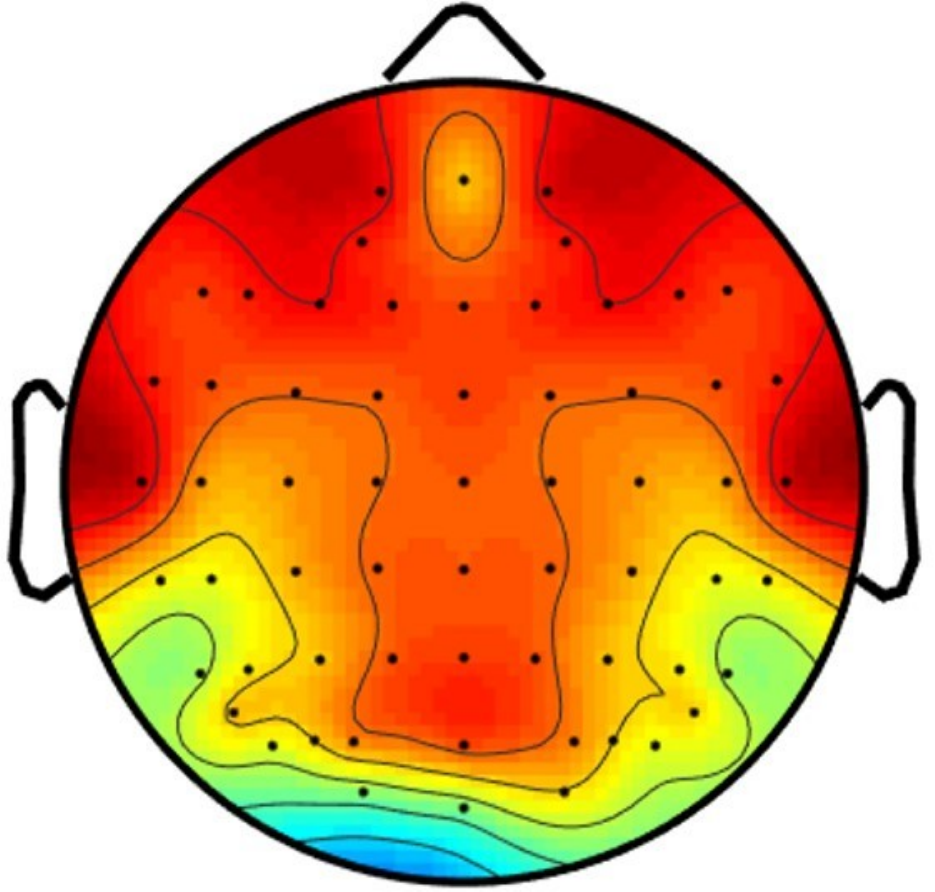}}
	\subfigure[Funny]{
		\label{Fig:MPED-dep-funny}
		\includegraphics[width=0.12\linewidth]{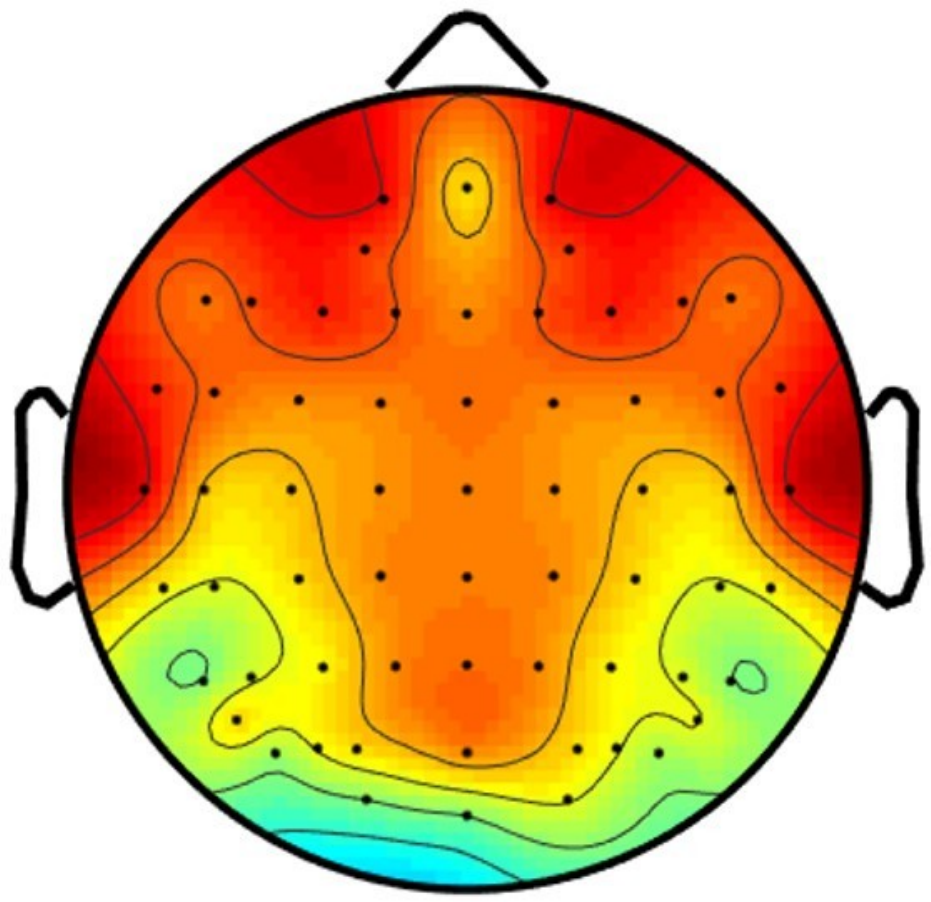}}
	\subfigure[Neutral]{
		\label{Fig:MPED-dep-neutral}
		\includegraphics[width=0.12\linewidth]{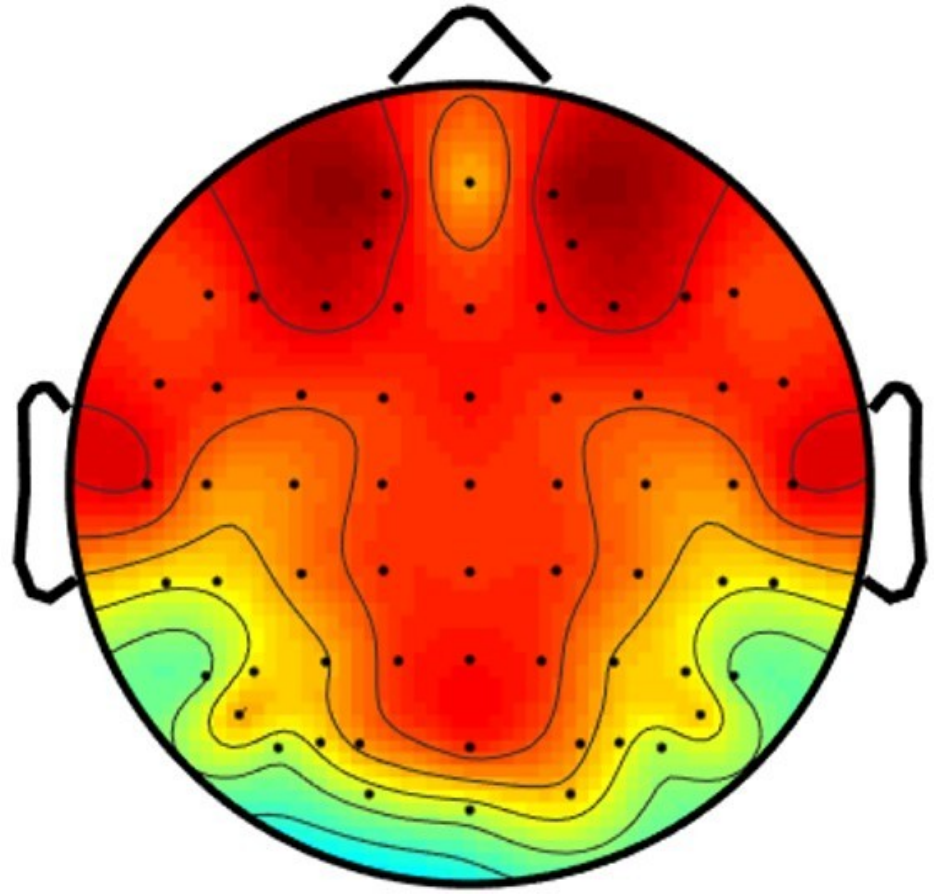}}	
	\subfigure[Sad]{
		\label{Fig:MPED-dep-sad}
		\includegraphics[width=0.12\linewidth]{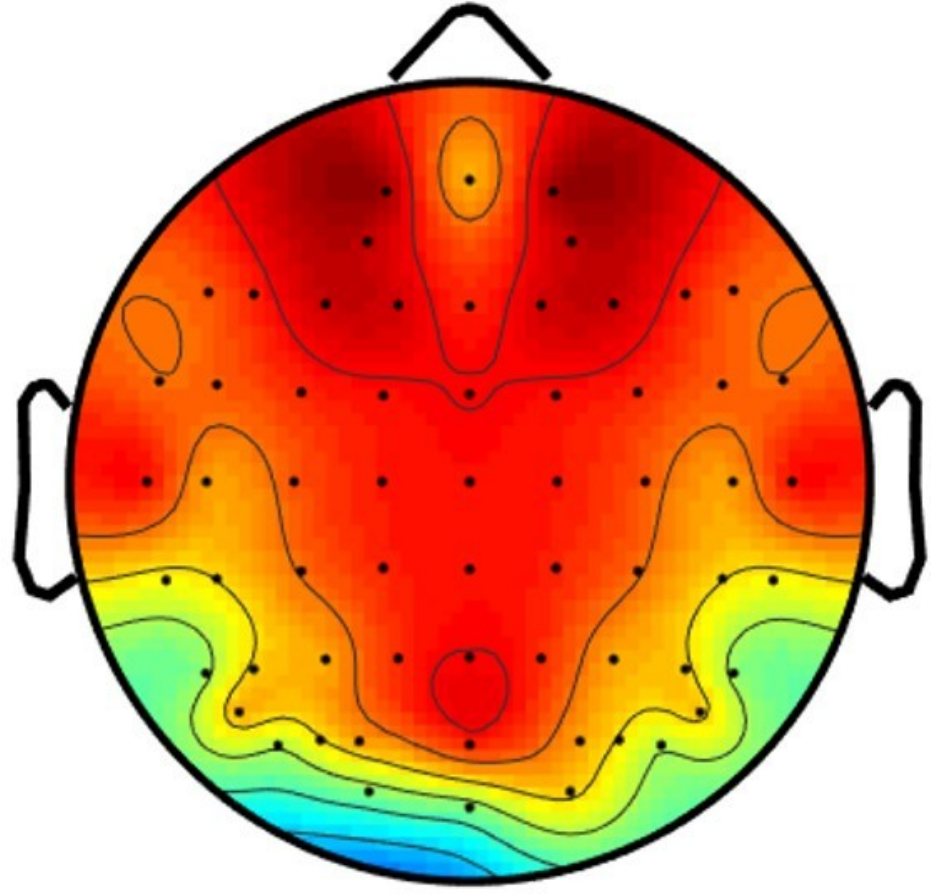}}
	\subfigure[Fear]{
		\label{Fig:MPED-dep-fear}
		\includegraphics[width=0.12\linewidth]{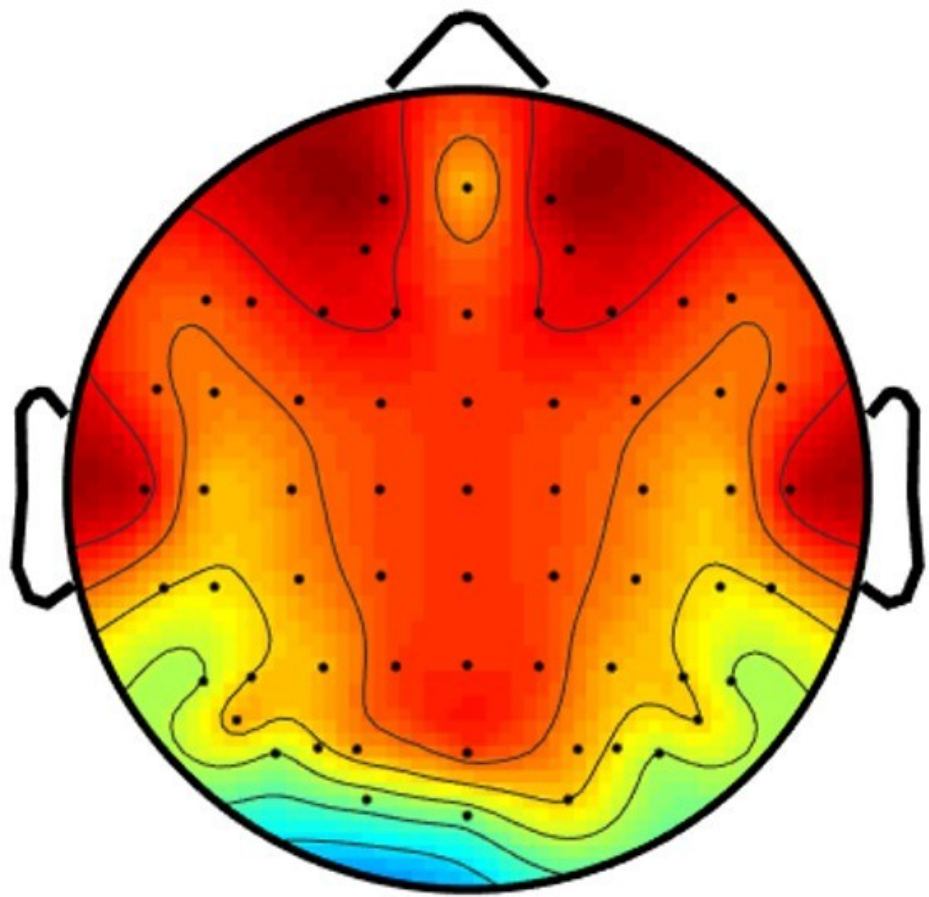}}
	\subfigure[Anger]{
		\label{Fig:MPED-dep-anger}
		\includegraphics[width=0.12\linewidth]{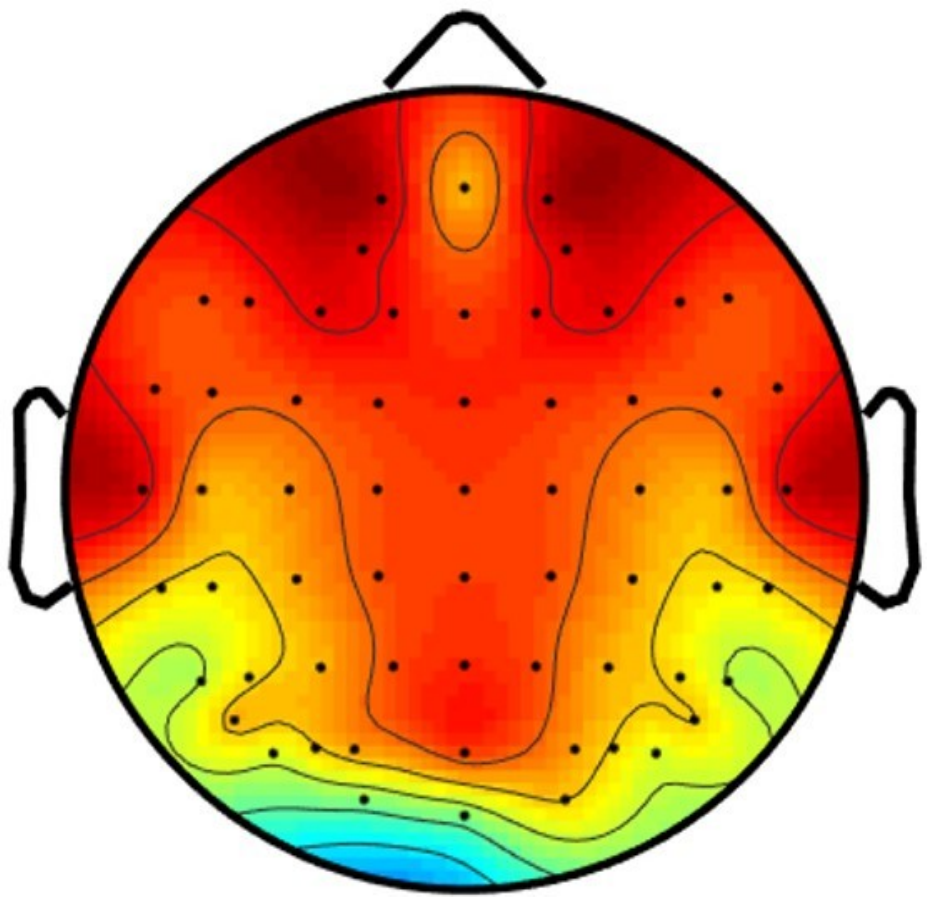}}
	\subfigure[Disgust]{
		\label{Fig:MPED-dep-disgust}
		\includegraphics[width=0.12\linewidth]{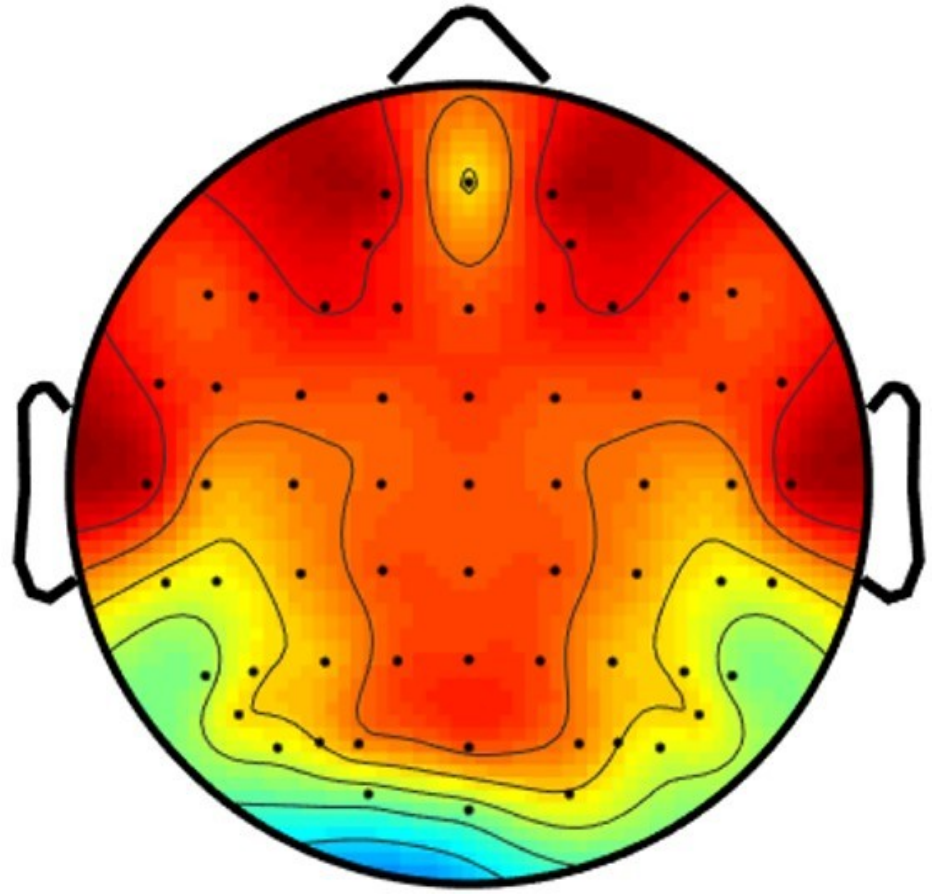}}
	\caption*{(3) MPED}	
	\caption{The EEG electrode activity maps in terms of different emotions based on the results of subject-dependent experiments. Darker red denotes more significant contribution. (a)-(c), (d)-(g), and (h)-(n) are the results on SEED, SEED-IV, and MPED datasets respectively. Note that these maps are symmetric because it shows the value computed by pairwise operation shared by paired electrodes. (Best Viewed in Color)}
	\label{Fig: Emotion activity maps}
\end{figure*}

Specifically, to explore where the differential information coming from in terms of the emotion expressed, we separately depict the electrode activity maps corresponding to each emotion in Fig.~\ref{Fig: Emotion activity maps}. Although it looks quite similar with Fig.~\ref{Fig: The EEG electrode activity maps}, i.e., the asymmetry on frontal and temporal lobes make more contribution to discriminate different emotions, we can observe some delicate distinctions from these maps of different emotions:
\begin{itemize}
\item [(1)] For the positive emotions (\textit{happy} in SEED and SEED-IV, \textit{joy} and \textit{funny} in MPED), we can see that the asymmetry on temporal lobe actives as same as (or even more than) the frontal lobe; 
\item [(2)] On the contrary, for the neutral emotion (Fig.~\ref{Fig:SEED-dep-neutral}, Fig.~\ref{Fig:SEED-IV-dep-neutral} and Fig.~\ref{Fig:MPED-dep-neutral}), the asymmetry on frontal lobe contributes more than temporal lobe;
\item [(3)] For the \textit{sad} emotion (Fig.~\ref{Fig:SEED-dep-sad}, Fig.~\ref{Fig:SEED-IV-dep-sad} and Fig.~\ref{Fig:MPED-dep-sad}), the asymmetry on frontal lobe basically dominates this emotion expression. But we can find that there is a broad activation area closing to the frontal-central lobe, especially in SEED-IV and MPED which contains more types of emotions than SEED;
\item [(4)] For the \textit{fear} emotion (Fig.~\ref{Fig:SEED-IV-dep-fear}), the asymmetry on frontal lobe dominates the emotion expression. But on MPED dataset, which includes more negative emotions (\textit{anger} and \textit{disgust}), the temporal lobe will have the same contribution with the frontal lobe.
\end{itemize}

\subsection{Electrodes reduction}
For the emotion recognition system in real-world applications, fewer electrodes will be preferred considering the feasibility and comfort. Thus in this section, we investigate how the performance varies with relatively small numbers of electrodes. Motivated by the results from Fig.~\ref{Fig: The EEG electrode activity maps} and~\ref{Fig: Emotion activity maps}, we select the paired electrodes on four brain areas referring to the locations of frontal and temporal lobes, denoted by \textit{Frontal (6)}, \textit{Frontal (10)}, \textit{Temporal (6)} and \textit{Temporal (9)} \footnote{Concretely, \textit{Frontal (6)} includes the paired electrodes of FP1-Fp2, AF3-AF4, F7-F8, F5-F6, F3-F4 and F1-F2; \textit{Frontal (10)} includes the paired electrodes of \textit{Frontal (6)} and FT7-FT8, FC5-FC6, FC3-FC4 and FC1-FC2; \textit{Temporal (6)} includes the paired electrodes of FT7-FT8, FC5-FC6, T7-T8, C5-C6, TP7-TP8 and CP5-CP6; \textit{Temporal (9)} includes the paired electrodes of \textit{Temporal (6)} and FC3-FC4, C3-C4 and CP3-CP4.}. The experimental results are shown in Table~\ref{Table: Electrodes reduction}. We can have two observation from this table:
\begin{itemize}
	 \item [(1)] The obtained recognition results based on fewer electrodes are comparable with that based on all the 31 paired electrodes especially on SEED dataset, which agrees with the observation of Fig.~\ref{Fig: The EEG electrode activity maps} and~\ref{Fig: Emotion activity maps}. It also verifies that the frontal and temporal lobes' asymmetry indeed contributes more to the EEG emotion recognition than the other brain areas. It is possible to consider utilizing fewer electrodes in EEG emotion recognition systems;
	 \item [(2)] Comparing between these two important brain regions, we can see the results based on temporal lobe electrodes outperform that based on frontal lobe. It seems temporal lobe associated more with emotion expression than frontal lobe for EEG emotion recognition;
\end{itemize}
\begin{table}[htb]
	\caption{The classification performance based on the frontal and temporal lobe EEG data for subject-dependent and subject-independent EEG emotion recognition on SEED, SEED-IV and MPED datasets.}
	\centering
	\renewcommand{\arraystretch}{1.3}
	\subtable[Subject-dependent experiment results]{\label{Table: Electrodes reduction - dep}
	\begin{threeparttable}		
		\begin{tabular}{|c|c|c|c|} 
			\hline
			\multirow{2}{*}{\textbf{Electrode area}} & \multicolumn{3}{c|}{\textbf{ACC / STD (\%)}}\\ \cline{2-4} 
			&  SEED            & SEED-IV          &   MPED  \\\hline
			Frontal (6)       & 80.15/09.86      & 57.93/13.88      & 29.02/05.68\\\hline
			Frontal (10)      & 84.49/08.83      & 63.02/16.95      & 32.37/06.79\\\hline
			Temporal (6)    & 88.16/08.03      & 64.88/15.76      & \textbf{33.61/07.19}\\\hline
			Temporal (9)    & \textbf{90.16/07.44}      & \textbf{65.19/16.03}      & 33.13/07.06\\\hline
			All (31)        &\textbf{93.12/06.06} &\textbf{74.35/14.09}&\textbf{40.34/07.59}\\\hline
		\end{tabular}
	\end{threeparttable}
	}
	\subtable[Subject-independent experiment results]{
	\begin{threeparttable}		
		\begin{tabular}{|c|c|c|c|c|c|c|c|} 
			\hline
			\multirow{2}{*}{\textbf{Electrode area}} & \multicolumn{3}{c|}{\textbf{ACC / STD (\%)}}\\ \cline{2-4} 
							&  SEED            & SEED-IV          &   MPED  \\\hline
			Frontal (6)     & 74.33/08.70& 67.28/08.19& 23.54/02.73\\\hline
			Frontal (10)    & 80.28/09.94& \textbf{68.16/07.85}& 25.44/04.95\\\hline
			Temporal (6)    & \textbf{85.04/07.13}& 65.07/08.74& 26.07/04.32\\\hline
			Temporal (9)    & 84.09/07.78& 66.92/08.74& \textbf{26.43/04.55}\\\hline
			All (31)        &\textbf{85.40/07.53}&\textbf{69.03/08.66}&\textbf{28.27/04.99}\\\hline
		\end{tabular}
		\begin{tablenotes}[para]
			\footnotesize 
		\end{tablenotes}
	\end{threeparttable}
	}
	\label{Table: Electrodes reduction}
\end{table}
\subsection{The performance based on single hemispheric EEG data}
From the above discussion, we can see the discrepancy information between the left and right hemispheres indeed contributes to the EEG emotion recognition task. On the other hand, it will be interesting to investigate which hemisphere is more tightly associated to emotion recognition. Therefore, in this section, we focus on this problem and conduct the same experiments by separately feeding our BiHDM with the left and right hemispheric data. The obtained experimental results are shown in Table~\ref{Table: Single hemisphere}, from which we can see the left hemisphere is superior to the right for EEG emotion recognition, especially in the experiments on SEED-IV and MPED datasets. Besides, comparing it with the results in Table~\ref{Table: Electrodes reduction - dep}, which are based on feeding the model with less symmetric electrodes' data, we can observe that results are comparable or even better than this experiment based on single hemisphere data. This verifies the effectiveness of discrepancy information for EEG emotion recognition from another aspect.
\begin{table}[htb]
	\caption{The classification performance based on single hemispheric EEG data for subject-dependent EEG emotion recognition on SEED, SEED-IV and MPED datasets.}
	\centering
	\renewcommand{\arraystretch}{1.3}
	\begin{threeparttable}		
		\begin{tabular}{|c|c|c|c|} 
			\hline
			\multirow{2}{*}{\textbf{Hemisphere}} & \multicolumn{3}{c|}{\textbf{ACC / STD (\%)}} \\ \cline{2-4}
			&  SEED            & SEED-IV          &   MPED\\ \hline
			BiHDM-left        & 86.63/08.88      & 64.48/15.34     & 35.92/07.26 \\ \hline
			BiHDM-right       & 86.39/07.54      & 60.11/13.53     & 33.08/08.30             \\ \hline
			BiHDM-overall    &\textbf{93.12/06.06} &\textbf{74.35/14.09}&\textbf{40.34/07.59}   \\ \hline
		\end{tabular}
		\begin{tablenotes}[para]
			\footnotesize 
		\end{tablenotes}
	\end{threeparttable}
	\label{Table: Single hemisphere}
\end{table}

\subsection{The effect of two directional RNNs to extract spatial information}
In BiHDM, the horizontal and vertical RNNs are adopted to model the structural relation between the electrodes. To evaluate the effect of this spatial information extraction for emotion recognition, we modified the framework of BiHDM with a single directional RNN, denoted by BiHDM-h and BiHDM-v respectively, to conduct the same experiments. The results are summarized in Table~\ref{Table: Two directional RNNs}, from which we can see that the predefined strategy of traversing the spatial region with horizontal and vertical RNNs achieves much better performance than the single directional RNN. This shows that the proposed spatial feature learning method is helpful to extract the discriminative information for EEG emotion recognition.
\begin{table}[htb]
	\caption{The classification performance of different spatial feature extraction methods for EEG emotion recognition on SEED, SEED-IV and MPED datasets.}
	\centering
	\renewcommand{\arraystretch}{1.3}
	\subtable[Subject-dependent experiment results]{
	\begin{threeparttable}		
		\begin{tabular}{|c|c|c|c|} 
			\hline
			\multirow{2}{*}{\textbf{Electrode area}} & \multicolumn{3}{c|}{\textbf{ACC / STD (\%)}}\\ \cline{2-4} 
			&  SEED            & SEED-IV          &   MPED  \\\hline
			BiHDM-h       & 87.47/09.17      & 62.06/15.01     & 36.24/08.41 \\\hline
			BiHDM-v       & 86.75/07.09      & 65.57/15.43     & 36.69/08.20 \\\hline
			BiHDM    &\textbf{93.12/06.06} &\textbf{74.35/14.09}&\textbf{40.34/07.59} \\\hline
		\end{tabular}
		\begin{tablenotes}[para]
			\footnotesize 
		\end{tablenotes}
	\end{threeparttable}
	}
	\subtable[Subject-independent experiment results]{
	\begin{threeparttable}		
		\begin{tabular}{|c|c|c|c|} 
			\hline
			\multirow{2}{*}{\textbf{Electrode area}} & \multicolumn{3}{c|}{\textbf{ACC / STD (\%)}}\\ \cline{2-4} 
							&  SEED              & SEED-IV            &   MPED  \\\hline
			BiHDM-h         & 82.38/09.33        & 66.80/08.22        & 28.05/04.98\\\hline
			BiHDM-v         & 81.03/10.28        & 66.96/08.28        & 27.86/05.06\\\hline
			BiHDM           &\textbf{85.40/07.53}&\textbf{69.03/08.66}&\textbf{28.27/04.99}\\\hline
		\end{tabular}
		\begin{tablenotes}[para]
			\footnotesize 
		\end{tablenotes}
	\end{threeparttable}
}
	\label{Table: Two directional RNNs}
\end{table}

\section{Conclusion}
\label{Sec: Conclusion}
In this paper, we propose a novel bi-hemispheric discrepancy model (BiHDM) for EEG emotion recognition. The proposed framework is easy to implement and generally achieves the state-of-the-art performance. This shows the effectiveness of incorporating the asymmetric differential information into EEG emotion recognition. In the future work, we will further investigate more left and right hemispheric differential operations to explore the potential efficacy of cerebral hemisphere asymmetry in EEG emotion recognition.

\ifCLASSOPTIONcaptionsoff
\newpage
\fi

\bibliographystyle{IEEEtran}
\bibliography{manuscript_bib}

\end{document}